\documentstyle[amssymb,epsfig]{mn2e}
\newcommand{\arc}{$^{\prime\prime}$}
\newcommand{\gtsim}{\mbox{{\raisebox{-0.4ex}{$\stackrel{>}{{\scriptstyle\sim}}
$}}}}
\newcommand{\ltsim}{\mbox{{\raisebox{-0.4ex}{$\stackrel{<}{{\scriptstyle\sim}}
$}}}}
\newcommand{\mc}{\multicolumn}
\def \kband{$K$-band }
\def \ukirt{United Kingdom Infrared Telescope (UKIRT)}
\def \uk{UKIRT}
\def \uf{UFTI}
\def \ui{UIST}
\def \gem{Gemini/NIRI}
\def \keck{Keck/NIRC}
\def \id{identification }
\def \rband{$R$-band }

\begin{document}

\title[The 6C** sample I: observational data]{The 6C** sample of steep-spectrum radio sources: I -- Radio data, near-infrared imaging and optical spectroscopy}

\author[Maria J.~Cruz et al.]
{Maria J.~Cruz$^{1,2}$\thanks{Email: mjc@astro.ox.ac.uk}, Matt J.~Jarvis$^{1}$,
  Katherine M.~Blundell$^{1}$, Steve Rawlings$^{1}$,
\and
 Steve Croft$^{3,4}$, Hans-Rainer Kl\"ockner$^{1}$, Ross J. McLure$^{5}$, Chris Simpson$^{6}$, \and Thomas A. Targett$^{5}$, Chris J.~Willott$^{7}$\\
\footnotesize
\\
$^{1}$Astrophysics, Department of Physics, Keble Road, Oxford, OX1 3RH, UK \\
$^{2}$Leiden University, Sterrewacht, Oort Gebouw, P.O. Box 9513, 2300 RA Leiden, The Netherlands\\
$^{3}$Institute of Geophysics and Planetary Physics, Lawrence Livermore National Laboratory, L-413, 7000 East Avenue, Livermore,\\ CA 94550 USA\\
$^{4}$ School of Natural Sciences, University of California Merced, P.O. Box
2039, Merced, CA 95344 USA\\
$^{5}$ Institute for Astronomy, University of Edimburgh, Royal
Observatory, Edimburgh, EH9 3HJ, UK\\
$^{6}$Astrophysics Research Institute, Liverpool John Moores University, Twelve
Quays House, Egerton Wharf, Birkenhead CH41 1LD, UK\\
$^{7}$Herzberg Institute of Astrophysics, National Research Council, 5071 West Saanich Road, Victoria, BC V9E 2E7, Canada
}

\maketitle

\begin{abstract}
We present basic observational data on the 6C** sample. This is a new
sample of radio sources drawn from the 151\,MHz 6C survey, which was
filtered with radio criteria chosen to optimize the chances of finding
radio galaxies at $z > 4$.  The filtering criteria are a
steep-spectral index and a small angular size. The final sample
consists of 68 sources from a region of sky covering 0.421\,sr.  We
present VLA radio maps, and the results of \kband imaging and optical
spectroscopy.

Near-infrared counterparts are identified for 66 of the 68 sources,
down to a $3 \sigma$ limiting magnitude of $K \sim 22$~mag in a
3-arcsec aperture. Eight of these identifications are spatially
compact, implying an unresolved nuclear source.  The $K$-magnitude
distribution peaks at a median $K \approx 18.7$~mag, and is found to
be statistically indistinguishable from that of the similarly selected
6C* sample, implying that the redshift distribution could extend to $z
\,\,\gtsim\,\, 4$.

Redshifts determined from spectroscopy are available for 22 (32 per
cent) of the sources, over the range of 0.2 $\,\,\ltsim\,\, z
\,\,\ltsim\,\, 3.3$ . We measure 15 of these, whereas the other 7 were
previously known. Six sources are at $z > 2.5$.  Four sources show
broad emission lines in their spectra and are classified as quasars.
Three of these show also an unresolved \kband identification.  Eleven
sources fail to show any distinctive emission and/or absorption
features in their spectra. We suggest that these could be (i) in the
so-called `redshift desert' region of $1.2 < z < 1.8$, or (ii) at a
greater redshift, but feature weak emission line spectra.

\end{abstract}

\section{INTRODUCTION}

The chief purpose of the 6C** sample was to find radio galaxies at the
highest redshifts ($z > 4$) and, based on its data, to directly
constrain the co-moving space density of high-redshift radio
sources. It is the second of the filtered 6C redshift surveys, being
preceded by a pilot sample, namely 6C* (Blundell et al. 1998; Jarvis
et al. 2001a,b).

\begin{table*}
\begin{center}
\begin{tabular}{lrrccr}
\hline
\mc{1}{c}{Sample} & \mc{1}{c}{Flux-density} & \mc{1}{c}{Sky Area} & \mc{1}{c}{Number of} & \mc{1}{c}{Spectroscopic} & \mc{1}{c}{Radio} \\
  & \mc{1}{c}{Limits (Jy)} & \mc{1}{c}{(sr)} & \mc{1}{c}{Sources} & \mc{1}{c}{Completeness} & \mc{1}{c}{Filtering}\\
 \hline
3CRR & $S_{178} \geq 10.9$ & 4.24 & 173 & 100\%& None\\
6CE  & $ 2.00 \leq S_{151} \leq 3.93$ & 0.10 & 58 & 97\% & None\\
7CRS & $S_{151} \geq 0.5$ & 0.022 & 128 & 90\% & None\\
\\
6C* & $ 0.96 \leq S_{151} \leq 2.0$ & 0.13 & 29 & 100\% & $\rm \alpha^{\scriptscriptstyle4.85{\scriptscriptstyle\rm G}}_{\scriptscriptstyle151{\scriptscriptstyle\rm M}} \geq 0.981$; $\theta<15$ arcsec\\
6C** & $S_{151} \geq 0.5$ & 0.42 & 68 & 32\% &  $\rm \alpha^{\scriptscriptstyle1.4{\scriptscriptstyle\rm G}}_{\scriptscriptstyle151{\scriptscriptstyle\rm M}} \geq 1.0$; $\theta<11$ arcsec\\
\hline
\end{tabular}
\end{center}
\caption{Summary of the properties of the 3CRR, 6CE, 7CRS, 6C*  and
  6C** redshift surveys. Note: spectroscopy has not been attempted yet
  on all members of the 6C** sample (see Section~\ref{sec:spectroscopy}). }
\label{tab:samples}
\end{table*}

Previous to the 6C and 7C surveys, low-frequency studies of radio
sources were based on the revised 3C sample (3CRR; Laing, Riley \&
Longair 1983).  The problem with this sample is that,
as with any single flux-limited sample, there is a tight correlation
between luminosity and redshift, in the sense that the highest
redshift sources in the sample are necessarily the most luminous ones.
One obvious implication is that based solely on the 3CRR sample, or in
any single flux-density limited sample, one cannot distinguish between
correlations of source parameters with redshift and with
luminosity. The other implication is that the 3CRR sample, being very
bright, does not find any objects at $z \gtrsim 2$, due to the rarity
of the extremely powerful (\mbox{$L_{151} \sim 10^{28}$ --
$10^{29}$~W~Hz$^{-1}$~sr$^{-1}$}) radio sources.

The 6CE (Rawlings, Eales \& Lacy 2001) and 7CRS (Willott et al. 1998,
2002; Lacy et al. 1999a,b) complete samples, with fainter flux-density
limits (see Table~\ref{tab:samples}), were designed to address these
problems and, in particular, to improve coverage of the 151\,MHz
luminosity ($L_{151}$), redshift plane. They were indeed very
successful in extending the $L_{151} - z$ plane into lower radio
luminosities at low-redshift and into $ 2 < z < 3$ at high radio
luminosity (e.g. Willott et al. 2001).  However, the problem with
these samples is that the requirement of low flux-density limits could
only be achieved, in terms of complete spectroscopic follow-up being
feasible, at the expense of the sky area covered.  This is
particularly dramatic in the case of the 7CRS sample and it very much
restricted the number of very high-redshift sources included in these
surveys. The need to overcome this limitation, i.e. to have a redshift
survey which covers a larger area of sky at flux levels comparable to
that of the 7CRS, and thereby probe very high redshift radio sources
($z > 3$) in a statistically meaningful way, provided the main
motivation for the construction of the filtered 6C redshift
surveys\footnote{A fainter and larger survey, containing $\sim$ 1000
radio sources with $S_{151} > 0.1$~Jy, is also being put together by
the groups at Oxford and Texas (Hill \& Rawlings 2003).  This survey,
named the TexOx-1000 (TOOT), aims to probe those radio-loud objects
which are more typical at high redshift ($z > 3$), i.e. less luminous
than the ones probed by 3CRR, 6CE and 7CRS at the same redshift, which
are rare.}.

These surveys, 6C* and 6C**, were both filtered with more selective
radio criteria than just a simple flux-density limit
(see Table~\ref{tab:samples}).
Additional criteria were introduced in order to filter out a large
fraction of the sources, mostly at low-redshift, thereby ensuring that
optical follow-up would be limited to a manageable number of sources,
biased towards high-redshift.
The chosen criteria took into account, therefore, the characteristics
often seen in very distant radio galaxies, namely steep radio spectral
index $\alpha$ and small angular size $\theta$.

The first criterion has been the one most widely used in searches for
the highest redshift radio galaxies (e.g. R\"ottgering et al. 1994;
Chambers et al. 1996; Blundell et al. 1998; De Breuck et al. 2000;
Cohen et al. 2004).
It is based primarily on the convex shape of the radio spectra which
flattens below rest-frame $\sim$ 300~MHz. Since higher redshift
objects are observed at higher rest-frame frequencies, their measured
spectral indices are invariably steeper. 
The second criterion relies on strong statistical
evidence for a negative evolutionary trend of linear size with
redshift (e.g. Barthel \& Miley 1988; Neeser et al. 1995; Blundell,
Rawlings \& Willott 1999)
although it has been
less widely used in targeting high-redshift radio sources.

Both criteria are imperfect in the sense that not all of the sources
at low-redshift are eliminated and some fraction at high redshift are
actually excluded. Jarvis et al. (2001b) quantified these fractions by
comparing the redshift distribution of the 6C* sample with the radio
luminosity function model of Willott et al. (2001).  They have found
that employment of the filtering criteria has reduced the number of $z
< 1.5$ sources by $\approx$ 90 per cent, whereas in the redshift range
$1.5 \,\,\ltsim\,\, z \, \,\ltsim \,\,3.0$ the fraction excluded
varies between $\sim 30 - 70$ per cent. The distribution, as a result,
is skewed towards objects at $z \gtrsim 2.0$, and the 6C* sample has a
median redshift of $z \sim 1.9$ (cf. 6CE and 7CRS with $\sim 1.1$
median redshift).

Even though the effects of the filtering criteria introduce
difficulties in interpreting the data from filtered samples, namely in
assessing which population of sources is filtered out, they are very
effective in biasing these samples to objects at high redshift. This
has been demonstrated very successfully with the 6C* sample (Jarvis et
al. 2001b), which led to the discovery of 6C0140+326 at $z = 4.41$
(Rawlings et al. 1996) -- the most distant radio galaxy known at the
time of discovery -- in a sample of only 29 objects.  Nevertheless,
the 6C* sample has only two objects at $z > 3$ and, therefore, the
$L_{151}-z$ diagram still suffers from small number statistics at high
redshift. The 6C** sample, deeper and larger than 6C*
(Table~\ref{tab:samples}), aims to ameliorate this problem and,
ultimately, to extend the $L_{151}-z$ diagram to even higher redshifts
($z \sim 5$).

The final spectroscopically complete 6C** sample will also potentially
supply us with targets with which we can probe the most massive
objects in the early Universe. Studies of the host galaxies of
radio-loud AGN have shown that all powerful radio sources are
associated with the most massive galaxies at every cosmological epoch
(Jarvis et al. 2001a; De Breuck et al. 2002; Willott et al. 2003; Zirm
et al. 2003).  Similarly, it has been shown that the most radio-loud
quasars are those with the most massive black-holes (e.g. Dunlop et
al. 2003; McLure \& Jarvis 2004).  Moreover, recent studies have
identified high redshift radio galaxies (HzRGs; $z > 2$) which are
associated with the sites of forming proto-clusters (Kurk et al. 2004;
Miley et al. 2004; Venemans et al. 2004, 2005), implying that HzRGs
are the most likely progenitors of the nearby brightest
clusters. Powerful radio sources are therefore key targets for studies
of the formation and evolution of massive structures in the early
Universe.  Also, the discovery of a radio galaxy within the epoch of
reionization ($z > 6$) will allow for 21~cm absorption line studies of
neutral hydrogen at that epoch (e.g. Carilli, Gnedin \& Owen 2002).

 Finally, one important characteristic of all the samples mentioned
 here is that they have been selected at low radio frequencies ($\nu <
 200$~MHz). This is a crucial factor in searches for high-redshift
 sources  (e.g. Blundell, Rawlings \& Willott 1999, 2000)
 and strongly determines the content of a sample.  Low-frequency
 surveys typically contain a high proportion of steep-spectrum
 sources, where the observed emission is dominated by the
 isotropically emitting radio-lobe components. In high-frequency ($\nu
 \gtrsim 1$~GHz) surveys, in contrast, core-dominated flat-spectrum
 sources prevail.
Consequently, surveys at high frequency tend to have more objects
which are selected above the flux limit because their emission is
boosted due to orientation of their jet axes, rather than their intrinsic
power. 
Surveys at low-frequency ($\nu < 200$~MHz),
being dominated by isotropic radio emission, provide us, therefore,
with much less biased samples of radio sources.

In this paper (hereafter Paper I) we present complete \kband imaging
of the 6C** sample, high-resolution radio maps for 42 of the sources,
and optical spectroscopy for a sub-sample of the optically identified
sources. In a companion paper (Paper II), we will present
single-colour ($K$-band) photometric redshifts for all the sources and
investigate the redshift distribution of the sample. The paper is set
out as follows. In Section~\ref{sec:sample} we describe the sample
selection criteria.  In Section~\ref{sec:observations} we outline the
observing and data reduction procedures. The radio maps, \kband images
and photometry, and the redshifts and line parameters are presented in
Section~\ref{sec:analysis}. In Section~\ref{sec:notes} we provide
notes of each source in the sample with respect to both imaging and
spectroscopy. The results are discussed in
Section~\ref{sec:discussion} and our main conclusions are summarized
in Section~\ref{sec:conclusions}.  Unless otherwise stated, we assume
throughout that $H_{0}=70~ {\rm km~s^{-1}Mpc^{-1}}$, $\Omega_ {\mathrm
M} = 0.3$ and $\Omega_ {\Lambda} = 0.7$. The convention used for radio
spectral index is $S_{\nu} \propto \nu^{-\alpha}$, where $S_{\nu}$ is
the flux-density at frequency $\nu$.

\begin{table*}
\scriptsize
\begin{center}
\begin{tabular}{lclrccrc}
\hline
 \mc{1}{c}{} & \mc{1}{c}{$K$-BAND}& \mc{1}{c}{} & \mc{1}{c}{} &
 \mc{1}{c}{RADIO} & \mc{1}{c}{} & \mc{1}{c}{} & \mc{1}{c}{} \\
\hline
\mc{1}{c}{Source} & \mc{1}{c}{Telescope/} & \mc{1}{c}{Dates Observed}
 & \mc{1}{c}{Total} & \mc{1}{c}{Map} & \mc{1}{c}{Dates} & \mc{1}{c}{Peak Flux} & \mc{1}{c}{RMS}\\
\mc{1}{c}{Name} & \mc{1}{c}{Instrument} & \mc{1}{c}{} &
 \mc{1}{c}{Exp. (s)} & \mc{1}{c}{Freq. (GHz)} & \mc{1}{c}{Obs.} & \mc{1}{c}{(mJy~beam$^{-1}$)} & \mc{1}{c}{(mJy~beam$^{-1}$)}\\
\hline
6C**0714+4616 & \uk/\uf & {\bf 2002-01-09} & 540 & 4.9  & 1996-12-06 & 22.16  & 0.36\\
6C**0717+5121 & \uk/\uf & {\bf 1999-03-12} & 3240& 8.4  & 1996-12-18 &~~1.30 & 0.10\\
            & \uk/\ui & 2003-01-28 & 540 \\
6C**0726+4938 & \uk/\ui & {\bf 2003-01-28} & 540 & 4.9  & 1996-12-18 &~~1.29 & 0.13\\
6C**0737+5618 & \uk/\uf & 1999-03-12 1999-03-13 & 4320\\
             & \uk/\ui & 2003-01-28 & 540\\
             & \gem   & {\bf 2004-02-03} & 2700& 8.4    & 1996-12-18 &~~0.35 & 0.09\\
\\
6C**0744+3702 & \uk/\uf & {\bf 1999-03-12} & 2160& 8.4  & 1996-12-18 &~~0.85 & 0.09\\
6C**0746+5445 & \uk/\uf & {\bf 1999-03-12} & 1620& 4.9  & 1996-12-06 &~~0.55 & 0.12\\
6C**0754+4640 & \uk/\uf & 1999-03-13 & 3000\\
            & \gem    & {\bf 2004-02-10} & 2460& 4.9    & 1996-12-06 &~~0.48 & 0.10\\
6C**0754+5019 & \uk/\ui & 2003-01-28 & 540\\
            & \gem    & {\bf 2004-02-10} & 2460& 4.9    & 1996-12-06 &~~8.75 & 0.21\\
\\
6C**0801+4903 & \uk/\uf & 1999-03-11$^{\dagger}$ {\bf 2003-12-18} & 1080 & 4.9 & 1996-12-06 &~~0.52 & 0.12\\
            & \uk/\ui & 2003-01-28 & 540\\
6C**0810+4605 & \uk/\uf & 2002-01-09 {\bf 2002-01-13} & 2160& 4.9 & 1996-12-06 &~~0.59 & 0.10\\
6C**0813+3725 & \uk/\uf & 2002-01-09 {\bf 2003-12-23} & 1080& 4.9 & 1996-12-06 & ~~0.34 & 0.08\\
6C**0824+5344 & \uk/\uf & 2002-01-09 {\bf 2002-01-13} & 2160& 4.9 & 1996-12-06 &~~5.56 & 0.19\\
6C**0829+3902 & \uk/\uf & {\bf 1999-03-13} & 2640 & 4.9 & 1996-12-06 &~~5.96 & 0.15\\
\\
6C**0832+4420 & \uk/\uf & 1999-03-11$^{\dagger}$ {\bf 2003-12-15} & 1080 & 4.9 & 1996-12-06 & ~~0.51 & 0.12\\
6C**0832+5443 & \uk/\uf & {\bf 2002-01-09} & 540& 4.9 & 1996-12-06 & ~~0.37 & 0.10\\
6C**0834+4129 & \uk/\uf & {\bf 2002-01-12} & 540& 8.4 & 1996-12-18 & ~~3.02 & 0.13\\
6C**0848+4803 & \uk/\uf & 1999-03-11$^{\dagger}$ {\bf 2003-12-15} & 1620& 4.9 & 1996-12-06 &~~3.68 & 0.12\\
6C**0848+4927 & \uk/\uf & {\bf 1999-03-13} & 1080& 4.9 & 1996-12-06 & 19.77 & 0.32\\
6C**0849+4658 & \uk/\uf & 1999-03-11$^{\dagger}$ {\bf 2003-12-15} & 1080& 8.4 & 1996-12-18 & 19.66 & 0.37\\
\\
6C**0854+3500 & \uk/\uf & {\bf 1999-03-12} & 540 & 8.4 & 1996-12-18 & ~~0.49 & 0.12\\
6C**0855+4428 & \uk/\uf & {\bf 2002-01-12} & 540 & 4.9 & 1996-12-06 & ~~3.33 & 0.15\\
6C**0856+4313 & \uk/\uf & {\bf 2003-12-15} & 540 & 4.9 & 1996-12-06 & ~~5.87 & 0.17\\
6C**0902+3827 & \uk/\uf & 2000-04-12 {\bf 2003-12-18} & 1080 & 8.4 & 1996-12-18 & ~~8.47 & 0.18\\
6C**0903+4251 & \uk/\uf & {\bf 2000-04-13} & 540 & 8.4 & 1996-12-18 & ~~8.23 & 0.18\\
6C**0909+4317 & \uk/\uf & {\bf 2000-04-12} & 3240& 4.9 & 1996-12-06 & ~~0.88 & 0.20\\
\\
6C**0912+3913 & \uk/\uf & {\bf 2000-04-13} & 540& 1.4 & & 26.03  & 0.37\\
6C**0920+5308 & \uk/\uf & {\bf 2002-11-22} & 540 & 4.9 & 1996-12-06 & ~~4.60 & 0.15\\
6C**0922+4216 & \uk/\uf & {\bf 2002-11-22} & 540 & 1.4 & & 199.65 & 0.13\\
6C**0924+4933 & \uk/\uf & {\bf 2002-11-22} & 540 & 4.9 & 1996-12-06 & ~~8.47 & 0.21\\
6C**0925+4155 & \uk/\uf & 2000-04-14 2003-12-18 2003-12-23 & 1620 \\
            & \gem    & 2004-02-12$^{\dagger}$ & 840\\
            & \keck   & {\bf 2004-04-30}  & 1920& 1.4 &  & 76.41 & 0.15\\         
\\
6C**0928+4203 & \uk/\uf & {\bf 2002-11-22} & 540 & 4.9 & 1996-12-06 & ~~8.52 & 0.24\\
6C**0928+5557 & \uk/\uf & {\bf 2000-04-14} & 540 & 4.9 & 1996-12-06 & ~~0.92 & 0.11\\
6C**0930+4856 & \uk/\uf & {\bf 2002-11-22} & 540 & 4.9 & 1996-12-18 & ~~0.79 & 0.19\\
6C**0935+4348 & \uk/\uf & 2000-04-12 2003-12-18 2003-12-23 & 540 \\
            & \gem    & {\bf 2004-02-12} & 2280& 1.4 &   & 29.02 & 0.14\\
6C**0935+5548 & \uk/\uf & {\bf 2000-04-13} & 540 & 4.9 & 1996-12-06 & ~~0.72 & 0.14\\
\\
6C**0938+3801 & \uk/\uf & 2000-04-13 {\bf 2004-02-09} & 1080 & 8.4 & 1996-12-18 & ~~2.89 & 0.13\\
            & \uk/\ui & 2003-01-28 & 540\\
6C**0943+4034 & \uk/\uf & {\bf 2000-04-14} & 540 & 1.4 & & 56.70 & 0.13\\
6C**0944+3946 & \uk/\uf & 2000-04-14 {\bf 2003-12-18} & 1080& 1.4 &  & 53.96 & 0.13\\
            & \uk/\ui & 2003-01-28 & 540\\
6C**0956+4735 & \uk/\uf & {\bf 2002-11-22} & 540 & 4.9 & 1996-12-06 & ~~0.75 & 0.18\\
\\
6C**0957+3955 & \uk/\uf & {\bf 2002-11-22} & 540& 8.4 & 1996-12-18 & ~~0.77 & 0.10\\
6C**1003+4827 & \uk/\uf & {\bf 2002-11-22} & 540& 4.9 & 1996-12-18 & ~~2.89 & 0.21\\
6C**1004+4531 & \uk/\uf & {\bf 2002-11-22} & 540& 4.9 & 1996-12-06 & ~~0.60 & 0.13\\
6C**1006+4135 & \uk/\uf & {\bf 2002-11-22} & 540& 8.4 & 1996-12-18 & ~~0.43 & 0.11\\
6C**1009+4327 & \uk/\uf & {\bf 2002-11-22} & 540& 1.4 &  & 133.03 & 0.14\\
6C**1015+5334 & \uk/\uf & {\bf 2000-04-14} & 540& 4.9 & 1996-12-06 & ~~0.58 & 0.12\\
\hline
\end{tabular}
{\caption{\label{tab:im_journalk1} Log of the \kband and radio imaging
observations of the sources present in the 6C$^{**}$ sample.  The bold
lettering indicates the \kband images shown in
Fig.~\ref{fig:kband_images1}. Columns 5, 6, 7 and 8 list the
frequency, observation date, peak flux density and rms noise of the
radio maps shown in Fig.~\ref{fig:kband_images1}. The maps at 4.9~GHz
and 8.4~GHz were obtained with the VLA in its A-configuration, as
described in Section~\ref{sec:radio}, with integrations times varying
from 2 to 5 minutes. The maps at 1.4~GHz are from the FIRST survey
(http://sundog.stsci.edu/first/catalogs.html). Note: the Keck--NIRC
image of 6C**0925+4155 was observed through a K$_{\rm S}$ filter.
$\dagger$ signifies that the observation was made under
non-photometric conditions.  }}
\end{center}
\end{table*}

\addtocounter{table}{-1}
\begin{table*}
\scriptsize
\begin{center}
\begin{tabular}{lclrccrc}
\hline
 \mc{1}{c}{} & \mc{1}{c}{$K$-BAND}& \mc{1}{c}{} & \mc{1}{c}{} &
 \mc{1}{c}{RADIO} & \mc{1}{c}{} & \mc{1}{c}{} & \mc{1}{c}{} \\
\hline
\mc{1}{c}{Source} & \mc{1}{c}{Telescope/} & \mc{1}{c}{Dates Observed}
 & \mc{1}{c}{Total} & \mc{1}{c}{Map} & \mc{1}{c}{Dates} & \mc{1}{c}{Peak Flux} & \mc{1}{c}{RMS}\\
\mc{1}{c}{Name} & \mc{1}{c}{Instrument} & \mc{1}{c}{} &
 \mc{1}{c}{Exp. (s)} & \mc{1}{c}{Freq. (GHz)} & \mc{1}{c}{Obs.} & \mc{1}{c}{(mJy~beam$^{-1}$)} & \mc{1}{c}{(mJy~beam$^{-1}$)}\\
\hline
6C**1017+3436 & \uk/\uf & {\bf 2000-04-14} & 540& 8.4 & 1996-12-18 & ~~0.48 & 0.12\\
6C**1018+4000 & \uk/\uf & {\bf 2000-04-14} & 540& 8.4 & 1996-12-18 & ~~0.44 & 0.11\\
6C**1035+4245 & \uk/\uf & {\bf 2000-04-12} & 3240& 8.4& 1996-12-18 & ~~0.44 & 0.11\\
6C**1036+4721 & \uk/\uf & {\bf 2002-11-22} & 540& 1.4 &  & 358.25 & 0.15\\
6C**1043+3714 & \uk/\uf & {\bf 2000-04-14} & 540& 1.4 &  & 191.04 & 0.30\\
6C**1044+4938 & \uk/\uf & 2000-04-13 {\bf 2000-04-14} & 3780& 8.4 & 1996-12-18 & ~~1.26 & 0.17\\
\\
6C**1045+4459 & \uk/\ui & {\bf 2003-01-28} & 540& 1.4 &  & 45.63 & 0.14\\
6C**1048+4434 & \uk/\ui & {\bf 2003-01-28} & 540& 1.4 &  & 86.02 & 0.14\\
6C**1050+5440 & \uk/\uf & {\bf 1999-03-06} & 3240& 1.4&  & 56.99 & 0.17\\
6C**1052+4349 & \uk/\uf & {\bf 2002-12-20} & 540 & 1.4&  & 45.77  & 0.13\\
6C**1056+3303 & \uk/\uf & {\bf 2002-12-20} & 540 & 1.4&  & 149.92 & 0.11\\
6C**1100+4417 & \uk/\uf & {\bf 2002-01-13} & 540 & 1.4&  & 59.79  & 0.14\\
\\
6C**1102+4329 & \uk/\uf & 2002-01-13 2003-12-18 {\bf 2003-12-23} & 1620& 1.4&  & 81.15 & 0.13\\
6C**1103+5352 & \uk/\uf & 2002-01-13 {\bf 2003-12-23} & 1080& 1.4&  & 256.49 & 0.14\\
6C**1105+4454 & \uk/\uf & {\bf 2002-01-13} & 540& 1.4&  & 36.60 & 0.14\\
6C**1106+5301 & \uk/\uf & {\bf 2002-05-05} & 540& 1.4&  & 36.27 & 0.14\\
6C**1112+4133 & \uk/\uf & {\bf 2002-12-20} & 540 & 1.4&  & 11.77 & 0.13\\
6C**1125+5548 & \uk/\uf & {\bf 2002-01-13} 2002-05-05 & 1080& 1.4&  & 24.44 & 0.14\\
\\
6C**1132+3209 & \uk/\uf & {\bf 2002-01-13} & 540& 1.4&   & 27.52 & 0.14\\
6C**1135+5122 & \uk/\uf & {\bf 2000-04-12} & 540 & 1.4&  & 50.91 & 0.14\\
6C**1138+3309 & \uk/\uf & 2002-01-13 2002-12-20 {\bf 2003-12-18}  & 1620& 1.4&  & 36.05 & 0.11\\
6C**1138+3803 & \uk/\uf & {\bf 2002-01-13} & 540& 1.4&  & 35.34 & 0.15\\
6C**1149+3509 & \uk/\uf & {\bf 2002-12-20} & 540 & 1.4&  & 54.16 & 0.13\\
\hline
\end{tabular}
{\caption{\label{tab:im_journal2} {\em continued}
}}
\end{center}
\end{table*}

\section{Sample Definition}\label{sec:sample}

The 6C** sample was drawn from Part II and Part III of the 6C survey
of radio sources at 151\,MHz (Hales et al. 1988, 1990).
The region of
Right Ascension between 07$^{\rm h}$\,14$^{\rm m}$ and 11$^{\rm
h}$\,50$^{\rm m}$ and Declination between +30$^{\circ}$ and
+58$^{\circ}$ was chosen and, in this area of the sky, all sources
with 151\,MHz flux density $S_{151}$ less than 0.5\,Jy were rejected.
Spectral indices were found by cross-correlating the remaining objects
with their counterparts (within 120~arcsec) in the NVSS catalogue at
1.4\,GHz  (Condon et al. 1998).
Any object with a spectral index between 151\,MHz and 1.4\,GHz (i.e.,
$\alpha^{\rm1400}_{\rm 151}$) that was flatter than 1.0 was then
rejected. From the remaining objects only those with angular size, as
listed in the NVSS catalogue, smaller than deconvolved 13 arcsec were
kept. Fulfilment of the above criteria resulted in a preliminary
sample of 158 sources. However, not all of these objects were
legitimate members of the 6C** sample.
 
One disadvantage of the 6C survey is its low spatial resolution
(4~arcmin). This means that in a sample like 6C**, with such a
relatively low flux-density limit, the problem of source confusion is
particularly acute. To ensure that sources with a  6C
(integrated) $S_{151}$ resulting from the confusion of two or more
radio sources -- which if considered individually would fall below the
flux-density limit -- were rejected, we cross-correlated the NVSS
(with 40~arcsec resolution) entries with those in the FIRST survey
(with 5~arcsec resolution) at  1.4\,GHz (Becker, White \& Helfand 1995).
Sources which were resolved into two or more independent radio sources
in the FIRST catalogue were excluded by eye. Through this procedure we
were also able to identify and exclude: (i) sources having an actual
angular size larger than the 13 arcsec limit, and (ii) sources which
were actually one of the hotspots of a much larger double source.

The final version of the 6C** sample excludes all these identified
`spurious' sources and
comprises 68 objects over an area of sky of 0.421\,sr.  We note also
that in this process some sources with sizes between 11 and 13 arcsec
were rejected, but not all of them. No sources smaller than
approximately 11 arcsec were rejected. Therefore, the sample is only
complete at an angular size limit of $\theta < 11$~arcsec.
 The final selection criteria may be summarized as follows:

\begin{enumerate}
\item $07^{\rm h}14^{\rm m} \leq$ R.A. (B1950) $\leq 11^{\rm
h}50^{\rm m}$, 
 
\item $30^\circ \leq$ Dec. (B1950) $\leq 58^\circ$,
 
\item $S_{151} \geq$ 0.5\,Jy,
 
\item $\rm \alpha^{1400}_{151} \geq 1.0$,

\item $\theta  < 13$~arcsec (complete at $\theta  < 11$~arcsec).

\end{enumerate}

In Table~\ref{tab:radiosurvey}, we list the 6C and NVSS positions and
fluxes, and the spectral indices derived from these, for all the
members of the 6C** sample.

\section{Observations and Data Reduction}\label{sec:observations}

\subsection{\kband imaging}\label{sec:imaging}

\subsubsection{UKIRT}\label{sec:UKIRT}

All the sources in the 6C** sample have been imaged in \kband
(2.2\,$\micron$) at the \ukirt. These observations were made over
several observing runs, starting in March 1999 up to February 2004,
using the UKIRT Fast-Track Imager (UFTI), except for one of the
runs. For the observations in January 2003, the UKIRT Imager
Spectrometer (UIST) was used. UFTI is a 1-2.5\,$\mu$m camera with a
1024 $\times$ 1024 HgCdTe array and a plate scale of 0.091 arcsec per
pixel, giving a field of view of \mbox{92\arc $\times$ 92\arc}. UIST
is a 1-5\,$\mu$m imager-spectrometer with a 1024 $\times$ 1024 InSb
array. In imaging mode there are two plate scales available, 0.12
arcsec per pixel or 0.06 arcsec per pixel. We used the former, giving
a field of view of 120\arc $\times$ 120\arc. All observations were
made in photometric conditions with the exception of the ones made on
1999 March 11 or unless otherwise stated, with seeing better than
0.6\,arcsec.

 The observational strategy used in all cases was the standard one
when observing with an infrared array. Each object field was observed
at nine (or a number multiple of nine) different positions in the
array, by offsetting the telescope by about 10 arcsec between each
exposure. The offsets were chosen so that they produced a $3 \times 3$
pointing pattern which allowed for sufficient spatial overlap between
the fields, while at the same time ensured that different elements of
the array sampled different parts of the field each time. This
technique is advantageous in that it provides for good flat-fielding
and sky subtraction, it gives a larger field-of-view and it makes it
easier to remove cosmic rays and bad pixels.  The integration time for
each of the multiple exposures was 60 seconds.  The total integration
time varied between objects in the sample depending on the magnitude
of the source and on telescope time constraints. However, most of the
sources were observed for at least 9 minutes or a multiple thereof.

\subsubsection{Gemini}

Five sources that remained undetected after the UKIRT campaign were
observed with the Near Infrared Imager (NIRI) on the Gemini North
Telescope in February 2004. NIRI comprises a 1024 $\times$ 1024
ALADDIN InSb array sensitive from 1 to 5\,$\mu$m, and three cameras
providing 0.022, 0.050, and 0.117 arcsec per pixel plate scales. We
used the lowest resolution camera, giving a field of view of 120\arc
$\times$ 120\arc. The sources were observed in \kband using the same
observational technique as with UKIRT. In a few cases, individual 60
second exposure frames were corrupted by electronic instability in the
detector and these were discarded.

\subsubsection{Keck}

One of the sources which was not detected with UKIRT was also observed
with the Near Infrared Camera (NIRC) on the Keck I Telescope in April
2004. The observations of this source with Gemini were lost due to bad
weather conditions. NIRC comprises a \mbox{256 $\times$ 256} InSb
array with a plate scale of 0.151 arcsec per pixel, giving a field of
view of \mbox{38.4\arc $\times$ 38.4\arc}. The source was observed
with the $K_{\rm S}$ filter for a total of 32 minutes, comprising 10
co-additions of 6 seconds per pointing in two 16-point dither
patterns. The seeing was 0.9\,arcsec.

\vskip 4.5mm 

A summary of all the \kband observations is given in
Table~\ref{tab:im_journalk1}.

\begin{table*}
\begin{center}
\scriptsize
\begin{tabular}{lrrrrcr}
\hline
\mc{1}{c}{Source Name} & \mc{2}{c}{Pointing Position}  & \mc{1}{c}{ Date} & \mc{1}{c}{Exposure} & \mc{1}{c}{Slit Width} & \mc{1}{c}{P.A.}  \\
\mc{1}{c}{} & \mc{2}{c}{(J2000)} & \mc{1}{c}{} & \mc{1}{c}{Time (s)} & \mc{1}{c}{(arcsec)} &  \mc{1}{c}{($^{\circ}$)}\\
\hline
6C**0714+4616 & 07 17 58.45 & 46 11 38.9 &  98Dec17 & 1x1200B, 1x1200R & 2.5 & 30\\
6C**0717+5121 &              &            & 98Dec18 & 1x1800B, 2x900R & 2.5 &  \\
              & 07 21 27.31 & 51 15 49.7 &  02Dec30 & 5x900B, 3x1800R & 1.0 & 234\\
6C**0726+4938 & 07 30 06.15 & 49 32 40.7 &  97Jan09 & 1x1800B, 2x900R & 2.5 & 135\\
6C**0737+5618 & 07 41 15.39 & 56 11 35.9 &  97Jan09 & 1x1800B, 2x900R & 2.5 & 28\\
              &             &            &  98Dec18 & 2x1800B, 4x900R & 2.5 &  \\
              & 07 41 15.38 & 56 11 35.8 &  98Dec19 & 2x1200B, 4x900R & 2.5 & 208\\
              & 07 41 15.39 & 56 11 35.9 &  98Dec20 & 3x1800B, 6x900R & 2.5 & 180\\
6C**0744+3702 & 07 47 29.36 & 36 54 38.0 &  97Jan09 & 2x1800B, 4x900R & 2.5 & 0\\
6C**0746+5445 &             &            &  98Dec18 & 1x1800B, 2x900R & 2.5 &  \\
              & {\bf 07 50 24.63} & {\bf 54 38 07.1} & {\bf 02Dec30} & {\bf 4x900B, 2x1800R} & {\bf 1.0} & {\bf 270}\\
6C**0754+4640 & 07 58 29.67 & 46 32 32.7 &  97Apr08 & 1x1200B, 2x900R & 2.5 & 14\\
6C**0754+5019 & 07 58 06.15 & 50 11 04.9 &  97Jan10 & 1x1800B, 2x900R  & 2.5& 33\\
6C**0810+4605 & 08 14 30.28 & 45 56 39.5 &  97Jan10 & 1x379B, 1x395R & 3.0 & 0\\
6C**0813+3725 & 08 16 53.72 & 37 15 54.8 &  97Jan10 & 1x1800B, 2x900R & 3.0 & 116\\
6C**0824+5344 & 08 27 58.91 &  53 34 15.0&  97Jan10 & 1x1200B, 2x600R & 2.5 & 137\\
6C**0829+3902 & 08 32 45.33 & 38 52 15.7 &  97Apr08 & 2x1200B, 2x900B & 2.5 & 60\\
              & 08 32 45.32 & 38 52 16.1 &  98Dec17 & 1x1800B, 2x900R & 2.5 & 60\\
6C**0832+5443 & 08 36 10.06 & 54 33 25.8 &  97Apr08 & 1x1500B, 2x750R & 2.5 & 83\\
              & {\bf 08 36 09.66} & {\bf 54 33 25.3}  & {\bf 98Dec17} & {\bf 1x1800B, 2x900R} & {\bf 2.5} & {\bf 83}\\
6C**0834+4129 & 08 37 49.23 & 41 19 53.0 & 98Dec17 & 1x1800B, 2x900R & 2.5 & 143\\
6C**0848+4927 & 08 52 14.84 & 49 15 44.7 & 98Dec20 & 1x1800B, 2x900R & 2.5 & 177\\
6C**0854+3500 & 08 57 15.98 & 34 48 24.1 & 02Dec31 & 2x900B, 1x1800R & 1.0 & 231\\
6C**0856+4313 &             &            & 98Dec18 & 1x1800B, 2x900R & 2.5 &  \\
6C**0902+3827 & 09 05 13.09 & 38 14 34.5  & 97Jan10 & 1x1800B, 2x900R & 2.5 & 0\\ 
6C**0925+4155 & 09 28 22.12 & 41 42 23.1  & 04Dec10 & 2x1200B, 2x1200R & 1.5 & 31\\
6C**0928+4203 & 09 31 38.55 & 41 49 45.4  & 98Dec17 & 1x1800B, 2x900R & 2.5 & 0\\
6C**0935+4348 & 09 38 21.41 & 43 34 37.4  & 04Dec10 & 2x1200B, 2x1200R & 1.5 & 90\\
6C**0938+3801 &             &             & 98Dec18 & 1x1800B, 2x900R & 2.5 &  \\
6C**1009+4327 & 10 12 09.89 & 43 13 09.1  & 05Apr14 & 1x1800B, 2x900R & 2.0 & 30\\
6C**1036+4721 & 10 39 15.67 & 47 05 40.3  & 03Jan01 & 1x900B, 1x900R  & 1.5 & 93\\
6C**1045+4459 & 10 48 32.25 & 44 44 28.3  &  97Jan09 & 1x1800B, 2x900R & 2.5 & 132\\
6C**1050+5440 & 10 53 36.30 & 54 24 41.9  &  97Jan09 & 1x1800B, 2x900R & 2.5 & 36\\
6C**1102+4329 & 11 05 43.00 & 43 13 24.6  &  05Apr14 & 1x1800B, 2x900R & 2.0 & 90\\
\hline
\end{tabular}
\end{center}
{\caption{\label{tab:spectra_journal} Log of the spectroscopic
observations obtained for sources in the 6C** sample. When there is
more than one observation for one source, the larger bold text
represents the observation for the spectra presented in
Fig.~\ref{fig:spectra}.  For the observations made in 1998 December 18
there is no record available of the pointing position nor of the slit
position angles. For this reason they do not appear in this table.}}
\end{table*}

\subsubsection{Data Reduction}

The near-infrared imaging data were reduced using a combination of
routines within IRAF. The data reduction comprised the following steps: (i)
dark-current subtraction; (ii) sky subtraction and flat-fielding,
using the multiple exposures of each field to create sky and
flat-field frames; and (iii) image co-adding. To combine the images
we adopted the following procedure.  First, a star common to all the
frames in a set of multiple exposures was identified.  Its centroid
pixel coordinates were then recorded in each frame and used to
calculate the relative shifts between frames (the offsets). When it
was not possible to identify a suitable object common to the frames,
we used the offsets calculated for another set of images in the same
night (we found this procedure to be more accurate than using the
image header offsets).  Each individual frame was scaled by a factor
of $10^{(0.4e X)}$, where $X$ is the airmass of the observation and $e$ is
the extinction coefficient given in magnitudes per airmass, in order
to correct for the small effects of atmospheric extinction at
$K$-band.
The individual images of each object field were then registered using
the offsets and combined into an image with intensity at each pixel
equal to the average of the pixels in the individual frames, which
excludes pixels more than 4$\sigma$ away from the median of the
distribution. The bad pixels were also excluded. These were flagged by
the means of a bad pixel mask and were determined 
from the flat fields.

\subsection{Radio Imaging}\label{sec:radio}

Most of the 6C** sources (42 out of 68) were observed with the Very
Large Array (VLA) in its A-configuration on various epoch during
December 1996. All observations were conducted using two snapshots
each typically of a few minutes in duration. For those candidates with
a sized quoted in the NVSS catalogue as resolved, the observations
were made at 1.4\,GHz, while those listed as unresolved were observed
at 5\,GHz. Data processing was done using standard procedures using
the AIPS software including self-calibration for phase corrections,
producing radio images with approximately 1\,arcsec and 0.35\,arcsec
resolution, respectively. For those images which were found to be
unresolved from these initial observations, we re-observed them in a
later epoch at 8\,GHz giving 0.2\,arcsec resolution.

The respective maps are presented in Fig.~\ref{fig:kband_images1}.
The frequencies used and other relevant data are listed in
Table~\ref{tab:im_journalk1}.  Where no high-resolution radio data
are available, we present maps from the FIRST survey.

\subsection{Optical Spectroscopy}\label{sec:spectroscopy}

The optical spectra presented in this paper were all obtained with the
ISIS long-slit spectrograph on the William Herschel Telescope (WHT).
The observations were made over the nights of 1997 January 9--10; 1997
April 8; 1998 December 17--20; 2002 December 30--31, 2003 January 1;
2004 December 10 and 2005 April 14. The observations on 2005 April 14
were part of the WHT service programme.  A journal of all these
observations is presented in Table~\ref{tab:spectra_journal}.

ISIS is a double-beam spectrograph which uses a dichroic to separate
the light into two different channels - the blue and red arms.  Each
of these is equipped with a detector optimized to operate at its
respective wavelength range.  All the observations were made using
both arms.  Observational setup details for each night are summarized
in Table~\ref{tab:obs_setup}.  Slit widths varied between 1 and 3
arcsec, depending on the seeing.  Conditions were photometric for all
nights except for the run in December 2002 / January 2003. The
atmospheric conditions in this run were not photometric since variable
cloud cover during each night affected the transparency.

Since all the sources observed were very faint, positioning and
centring of the long slit on targets were made by off-setting from a
nearby bright star.  In the first three runs (January 1997, April 1997
and December 1998), since most objects had not been previously
identified through optical imaging, most spectra were taken by
``blind'' offsetting to the radio position and by aligning the slit
with the radio axis (Rawlings, Eales \& Warren 1990).
For the observations made after 2002 all of the spectra, except for
6C**0935+4348, were taken at the near-infrared counterpart position.

\begin{table*}
\begin{center}
\scriptsize
\begin{tabular}{l|l|lllll}
\hline
\mc{2}{c|}{Dates} & \mc{1}{c}{1997 Jan 9-10} & \mc{1}{c}{1998 Dec 17-20} & \mc{1}{c}{2002 Dec 30-31} & \mc{1}{c}{2004 Dec 10} & \mc{1}{c}{2005 Apr 14}\\
\mc{2}{c|}{}      & \mc{1}{c}{1997 Apr 8} &                              & \mc{1}{c}{2003 Jan 01} & \mc{1}{c}{} &                       \\
\hline
\mc{2}{c|}{Dichroic} & \mc{1}{l}{6100} & \mc{1}{l}{6100} & \mc{1}{l}{5400} & \mc{1}{l}{5400} & \mc{1}{l}{5300}\\
\hline
     & CCD            & TEK5          &  EEV12         & EEV12          & EEV12          & EEV12\\
     & CCD size       & 1124x400      & 720x4200       & 401x4200       & 1201x4200      & 601x2100\\
     & Binning factor &  1x1          &  1x1           & 1x1            & 1x1            & 2x2\\
Blue & Binned pixel   & 0.36 arcsec      & 0.19 arcsec       & 0.19 arcsec       & 0.19 arcsec       & 0.38 arcsec\\
     & Grating        & R158B         & R158B          & R158B / R300B    & R300B          & R158B\\
     & Spectral scale & 2.9 \AA\,pixel$^{-1}$ & 1.62 \AA\,pixel$^{-1}$ & 1.62 / 0.86 \AA\,pixel$^{-1}$ & 0.86 \AA\,pixel$^{-1}$ & 3.2 \AA\,pixel$^{-1}$\\
     & Spectral range & 2970 \AA      & 3539 \AA       & 5670 / 3024 \AA  & 3024 \AA       & 3427 \AA\\
\hline
     & CCD            & TEK2          &  TEK2         & MARCONI2       & MARCONI2       & MARCONI2\\
     & CCD size       & 1124x400      & 1124x400      & 401x4700       & 1201x4700      & 501x2350\\
     & Binning        &  1x1          &  1x1          & 1x1            & 1x1            & 2x2\\
Red  & Binned pixel   & 0.36 arcsec      & 0.36 arcsec      & 0.19 arcsec       & 0.19 arcsec       & 0.38 arcsec\\
     & Grating        & R158R         & R158R         & R158R / R316R  & R316R          & R158R\\
     & Spectral scale & 2.9 \AA\,pixel$^{-1}$ & 2.9 \AA\,pixel$^{-1}$ & 1.63 / 0.84 \AA\,pixel$^{-1}$ & 0.84 \AA\,pixel$^{-1}$ & 3.3 \AA\,pixel$^{-1}$\\
     & Spectral range & 2970 \AA      & 2970 \AA      & 4492 / 2302 \AA  & 2302 \AA       & 4637 \AA\\
\hline
\end{tabular}
\end{center}
{\caption{\label{tab:obs_setup} Summary of the observational setup for
    each optical spectroscopy run.}}
\end{table*}

\subsubsection{Data Reduction}
\label{sec:spectra-datareduction}

Reduction of the raw spectra was carried out using standard routines
within the {\scriptsize IRAF NOAO} packages, particularly the
{\scriptsize TWODSPEC} and {\scriptsize ONEDSPEC} packages. Data
reduction involved the following steps, which were performed separately
and in parallel for the red and blue arms: (i) overscan correction and
bias subtraction; (ii) flat-fielding, using Tungsten lamp spectra in
conjunction with twilight sky spectra; (iii) wavelength calibration
and geometrical rectification, using comparison arc lamps and standard
stars, respectively; (iv) background subtraction; and, (v) extinction
correction. Spectrophotometric standard stars were observed during the
course of each night and were used to flux-calibrate the data.

One-dimensional spectra were extracted from the fully-reduced
two-dimensional spectra. The extractions were made over an aperture
equal to the full-width at zero-intensity (FWZI), making sure that
extended emission lines were fully enclosed by it.  When more than one
exposure of the same object was taken using the same observational
setup and slit P.A., the individual one-dimensional spectra were
combined using a high-threshold rejection to eliminate cosmic
rays. The value of the threshold was set to be slightly higher than
the maximum pixel count due to real features in the spectra.

Reduced one-dimensional spectra for all the objects that have obvious
emission lines are presented in Fig.~\ref{fig:spectra}. For
presentation purposes all have been boxcar smoothed over three
pixels. Cosmic rays were edited out of the one-dimensional spectra by
inspection of the two-dimensional spectra. If features appear in only
one arm of the spectrum then the other is not shown. Otherwise, the
spectra from the red and blue arms were joined together by averaging
over $\approx$~50~\AA \, in the overlapping regions.

\section{Analysis}\label{sec:analysis}

\subsection{Source Identification and \kband Photometry}\label{sec:photometry}

\subsubsection{Astrometry and identification}
Astrometry for each image was achieved by comparing the image with a
finding chart from the Second Palomar Observatory Sky Survey.  For
most fields three or more sources were common to both the image and
the finding chart. For these cases the plate solution was obtained
using the KARMA--{\small KOORDS} task package (Gooch 1996).  For cases
where fewer than three POSS-II sources were detected on the K-band
image, the plate solution for another image with a good fit on the
same observing run was used along with the position of one of the
detected stars to fix the astrometry.  After this procedure, and where
possible, we compared star positions in our fields with those from the
USNO-A2.0 Catalogue (Monet et al. 1998) and the 2MASS Catalogue
(Skrutskie et al. 2006). For most cases we found differences in the
range of 0.05 to 0.5 arcsec. Where necessary (thirteen sources) we
used stars from these two catalogues to improve the astrometric
solution.

Identification of the
near-infrared counterparts to the radio sources was accomplished by
overlaying the radio contour maps on to the \kband images using the
KARMA--{\small KVIS} task package (Gooch 1996). 
The reduced and astrometrized \kband images, with radio contours
overlaid, are shown in Fig.~\ref{fig:kband_images1}. All the sources
are discussed individually in Section~\ref{sec:notes-imaging}.

\subsubsection{Photometry}
We measured magnitudes for all radio source identifications using the
{\scriptsize IRAF} task {\scriptsize PHOT} and circular apertures with
diameters of 3, 5 and 8 arcsec. This is to facilitate comparison with
previous work and to make sure that a magnitude is obtained even when
there is a nearby object. In Table~\ref{tab:mag_journalk1} we present
the \kband magnitudes for all of the sources in the 6C** sample.

The photometry was calibrated using observations of standard stars
selected from the UKIRT list of faint standards (Hawarden et
al. 2001).  Two to three different stars were observed over the course
of each night with 5 to 20 second exposures (depending on their
magnitudes) in mosaics of five positions.  The reduction process for
these observations is identical to the one outlined in
Section~\ref{sec:imaging}.  The following zero-points were obtained:
$20.81 \pm 0.02$~mag for the single night of observations with UIST;
in the range of \mbox{21.95 -- 22.36~mag} for the several nights where
UFTI was used (the values remained fairly constant throughout
individual nights in a given observing run but changed significantly
between different runs); and $\approx 23.6$ mag for all the
observations with NIRI. For the observations made with NIRC at Keck
the $K_{\rm S}$ zero-point was 24.66~mag. We used this zero-point to
measure the $K_{\rm S}$ magnitude for 6C**0925+4155 (the source
observed with NIRC).  We do not convert this $K_{S}$ magnitude to
$K$-band as the correction factor is smaller than the photometric
uncertainty.  In the remainder of this paper we will refer to it as a
$K$-band magnitude.

 \begin{table*}
 \scriptsize
 \begin{center}
 \begin{tabular}{lccllll}
 \hline
 \mc{1}{c}{Source Name} & \mc{2}{c}{NIR Position (J2000)} & \mc{1}{c}{Magnitude} & \mc{1}{c}{Magnitude} & \mc{1}{c}{Magnitude} & \mc{1}{c}{Detector}\\
 \mc{1}{c}{} & \mc{1}{c}{RA} & \mc{1}{c}{DEC} & \mc{1}{c}{3'' diameter}       & \mc{1}{c}{5'' diameter}       & \mc{1}{c}{8'' diameter} & \\
 \hline
 6C**0714+4616$^{\dag}$ & 07 17 58.48 & +46 11 39.3 & 16.705 $\pm$ 0.022 & 16.565 $\pm$ 0.039 & 16.316 $\pm$ 0.069 & \uf\\
 6C**0717+5121 & 07 21 27.23 & +51 15 50.3 & 18.148 $\pm$ 0.033 & 17.788 $\pm$ 0.047 & nbo                & \uf \\
             &              &             & 18.081 $\pm$ 0.070 & 17.804 $\pm$ 0.098 & nbo                & \ui \\
 6C**0726+4938 & 07 30 06.20 & +49 32 41.2 & 18.991 $\pm$ 0.220 & 18.711 $\pm$ 0.395 & 18.168 $\pm$ 0.497 & \ui \\
 6C**0737+5618 &     --       &      --      & $> 21.8$           & $>21.5$            &  $> 21.1$          & NIRI\\
 \\
 6C**0744+3702 & 07 47 29.36 & +36 54 37.9 & 19.708 $\pm$ 0.142 & 19.667 $\pm$ 0.220 & 19.440 $\pm$ 0.753 & \uf \\
 6C**0746+5445 & 07 50 24.63 & +54 38 06.9 & 18.423 $\pm$ 0.061 & nbo                & nbo                & \uf \\
 6C**0754+4640(a) & 07 58 29.70 & +46 32 34.0 & 20.138 $\pm$ 0.049 & 19.971 $\pm$ 0.082 & nbo                & NIRI\\
6C**0754+4640(b)  & 07 58 29.43 & +46 32 30.8 & 20.162 $\pm$ 0.051 & 19.907 $\pm$ 0.079 & nbo          & NIRI\\
 6C**0754+5019 & 07 58 06.06 & +50 11 03.1 & 21.519 $\pm$ 0.186 & 20.629 $\pm$ 0.265 & nbo                & NIRI\\  
\\
 6C**0801+4903 & 08 04 41.26 & +48 54 58.3 & 19.979 $\pm$ 0.433 & 19.855 $\pm$ 1.529 & nbo                & \uf \\
 6C**0810+4605 & 08 14 30.31 & +45 56 39.8 & 16.405 $\pm$ 0.010 & 16.123 $\pm$ 0.016 & 15.993 $\pm$ 0.030 & \uf \\ 
 6C**0813+3725 & 08 16 53.70 & +37 15 54.8 & 20.018 $\pm$ 0.450 & 19.574 $\pm$ 0.540 & 18.798 $\pm$ 0.684 & \uf \\ 
 6C**0824+5344 & 08 27 58.87 & +53 34 15.4 & 19.720 $\pm$ 0.231 & 19.719 $\pm$ 0.328 & 19.392 $\pm$ 0.721 & \uf \\  
 6C**0829+3902 & 08 32 45.32 & +38 52 16.7 & 19.427 $\pm$ 0.108 & 19.413 $\pm$ 0.227 & nbo                & \uf \\ 
 \\
 6C**0832+4420 & 08 35 27.84 & +44 09 52.4 & 18.896 $\pm$ 0.105 & 19.149 $\pm$ 0.266 & 18.915 $\pm$ 0.406 & \uf \\
 6C**0832+5443 & 08 36 09.87 & +54 33 25.8 & 19.604 $\pm$ 0.230 & 19.283 $\pm$ 0.441 & nbo                & \uf \\
 6C**0834+4129 & 08 37 49.25 & +41 19 54.6 & 19.571 $\pm$ 0.192 & 19.396 $\pm$ 0.460 & 19.378 $\pm$ 0.717 & \uf \\
 6C**0848+4803 & 08 52 17.86 & +47 52 21.4 & 18.581 $\pm$ 0.077 & 18.169 $\pm$ 0.108 & 17.828 $\pm$ 0.161 & \uf \\
 6C**0848+4927 & 08 52 14.83 & +49 15 44.8 & 18.577 $\pm$ 0.078 & 18.354 $\pm$ 0.134 & 18.222 $\pm$ 0.263 & \uf \\
 \\
 6C**0849+4658$^{\dag}$ & 08 53 09.47 & +46 47 00.9 & 17.689 $\pm$ 0.037 & 17.500 $\pm$ 0.058 & 17.319 $\pm$ 0.099 & \uf \\
 6C**0854+3500 & 08 57 15.99 & +34 48 24.9 & 18.388 $\pm$ 0.101 & 18.273 $\pm$ 0.141 & 18.121 $\pm$ 0.177 & \uf \\
 6C**0855+4428 & 08 58 38.57 & +44 16 26.6 & 18.315 $\pm$ 0.086 & 18.485 $\pm$ 0.202 & nbo                & \uf \\
 6C**0856+4313 & 08 59 20.19 & +43 02 00.8 & 18.568 $\pm$ 0.072 & 18.234 $\pm$ 0.094 & 17.999 $\pm$ 0.119 & \uf \\
 6C**0902+3827 & 09 05 13.10 & +38 14 34.5 & 19.544 $\pm$ 0.224 & 19.290 $\pm$ 0.358 & nbo                & \uf \\
 \\
 6C**0903+4251 & 09 06 26.15 & +42 39 04.2 & 16.947 $\pm$ 0.029 & 16.648 $\pm$ 0.046 & 16.615 $\pm$ 0.090 & \uf \\
 6C**0909+4317 & 09 13 00.80  & +43 05 20.1 & 18.977 $\pm$ 0.066 & 18.784 $\pm$ 0.123 & 18.635 $\pm$ 0.235 & \uf \\
 6C**0912+3913 & 09 16 05.11  & +39 00 19.3 & 18.695 $\pm$ 0.146 & 18.032 $\pm$ 0.180 & 17.595 $\pm$ 0.304 & \uf \\
 6C**0920+5308 & 09 23 47.60  & +52 56 44.8 & 14.982 $\pm$ 0.004 & 14.707 $\pm$ 0.006 & 14.526 $\pm$ 0.009 & \uf \\
 6C**0922+4216$^{\dag}$ & 09 25 59.28  & +42 03 37.7 & 15.035 $\pm$ 0.005 & 15.961 $\pm$ 0.008 & 15.928 $\pm$ 0.015 & \uf \\
\\
 6C**0924+4933 & 09 27 55.66  & +49 21 16.3 & 15.446 $\pm$ 0.007 & 15.126 $\pm$ 0.009 & 14.955 $\pm$ 0.015 & \uf \\
6C**0925+4155$^{\ddag}$ & 09 28 22.18  & +41 42 24.9 & 20.080 $\pm$ 0.078 & 20.279 $\pm$ 0.192 & nbo & NIRC\\
 6C**0928+4203 & 09 31 38.56  & +41 49 44.8 & 18.691 $\pm$ 0.129 & 18.661 $\pm$ 0.256 & 18.448 $\pm$ 0.403 & \uf \\
 6C**0928+5557 & 09 32 17.64  & +55 44 42.5 & 17.824 $\pm$ 0.055 & 17.285 $\pm$ 0.064 & nbo                & \uf \\
 6C**0930+4856 & 09 34 14.55  & +48 42 46.0 & 18.843 $\pm$ 0.215 & 18.950 $\pm$ 0.584 & 18.903 $\pm$ 1.164 & \uf \\
\\
 6C**0935+4348 &    --         &   --        & $>21.7$            & $>21.6$             & $>20.9$            & NIRI\\
  6C**0935+5548 & 09 39 04.73 & +55 35 09.0 & 18.951 $\pm$ 0.390 & 18.649 $\pm$ 0.273 & 18.325 $\pm$ 0.348 & \uf \\
 6C**0938+3801 & 09 41 52.27 & +37 47 24.9 & 19.244 $\pm$ 0.318 & 18.552 $\pm$ 0.279 & 18.132 $\pm$ 0.338 & \uf \\
             &              &             & 18.946 $\pm$ 0.081 & 18.353 $\pm$ 0.100  & 18.038 $\pm$ 0.160 & \uf \\
6C**0943+4034 & 09 46 27.45 & +40 20 33.1 & 18.098 $\pm$ 0.069 & 17.724 $\pm$ 0.100 & 17.592 $\pm$ 0.199 & \uf \\
 \hline
 \end{tabular}
 \end{center}
 {\caption{\label{tab:mag_journalk1} The \kband magnitudes for the 6C** sample
in three different apertures. Columns 2--3 list the position of the near-infrared identification for each source. In columns 4--6 `nbo' denotes that the radio galaxy is too close to a nearby object to measure the magnitude reliably. 
Note: $^{\ddag}$6C**0925+4155 was observed through a $K_{S}$
filter and therefore the magnitude presented here is a $K_{S}$
magnitude (see also the text in Section~\ref{sec:photometry}). $\dag$
signifies that the source has an unresolved near-infrared identification. 
}}
 \end{table*}

 \addtocounter{table}{-1}
\begin{table*}
 \scriptsize
 \begin{center}
 \begin{tabular}{lccllll}
 \hline
 \mc{1}{c}{Source Name} & \mc{2}{c}{NIR Position (J2000)} & \mc{1}{c}{Magnitude from} & \mc{1}{c}{Magnitude from} & \mc{1}{c}{Magnitude from} & \mc{1}{c}{Detector}\\
 \mc{1}{c}{}   & \mc{1}{c}{RA} & \mc{1}{c}{DEC} & \mc{1}{c}{3'' diameter}       & \mc{1}{c}{5'' diameter}       & \mc{1}{c}{8'' diameter} & \\
 \hline  
 6C**0944+3946 & 09 47 49.16 & +39 33 10.4 & 19.605 $\pm$ 0.214 & 18.938 $\pm$ 0.399 & 19.088 $\pm$ 0.981 & \uf \\
 6C**0956+4735 & 09 59 18.81 & +47 21 13.4 & 17.799 $\pm$ 0.049 & 17.458 $\pm$ 0.063 & 17.192 $\pm$ 0.090 & \uf \\
 6C**0957+3955 & 10 00 46.12 & +39 40 45.5 & 18.571 $\pm$ 0.107 & 18.462 $\pm$ 0.187 & 18.264 $\pm$ 0.317 & \uf \\
 6C**1003+4827$^{\dag}$ & 10 06 40.55 & +48 13 09.6 & 17.304 $\pm$ 0.033 & 17.092 $\pm$ 0.052 & 16.950 $\pm$ 0.092 & \uf \\
 6C**1004+4531 & 10 07 42.82 & +45 16 07.6 & 17.998 $\pm$ 0.061 & 17.539 $\pm$ 0.080 & 17.183 $\pm$ 0.115 & \uf \\
\\
 6C**1006+4135 & 10 09 27.47 & +41 20 46.3 & 19.268 $\pm$ 0.167 & 19.825 $\pm$ 0.433 & nbo                & \uf \\
 6C**1009+4327 & 10 12 09.88 & +43 13 09.0 & 20.513 $\pm$ 0.596 & nbo                & nbo                & \uf \\
 6C**1015+5334 & 10 18 29.93 & +53 19 33.6 & 19.159 $\pm$ 0.188 & 18.797 $\pm$ 0.289 & 18.516 $\pm$ 0.595 & \uf \\
 6C**1017+3436 & 10 20 05.70 & +34 21 19.8 & 19.314 $\pm$ 0.185 & 18.999 $\pm$ 0.247 & 18.972 $\pm$ 0.389 & \uf \\
 6C**1018+4000 & 10 21 28.70 & +39 45 45.3 & 18.568 $\pm$ 0.097 & 18.245 $\pm$ 0.130 & 18.434 $\pm$ 0.258 & \uf \\
 \\
 6C**1035+4245 & 10 38 41.03 & +42 29 51.9 & 17.463 $\pm$ 0.016 & 17.269 $\pm $ 0.024 & 17.250 $\pm$ 0.048 & \uf \\
 6C**1036+4721$^{\dag}$ & 10 39 15.67 & +47 05 40.5 & 17.099 $\pm$ 0.025 & 17.100 $\pm $ 0.043 & 16.967 $\pm$ 0.065 & \uf \\
 6C**1043+3714 & 10 46 11.91 & +36 58 44.5 & 17.579 $\pm$ 0.016 & nbo                & nbo                & \uf \\
 6C**1044+4938 & 10 47 47.86 & +49 22 36.1 & 19.096 $\pm$ 0.319 & 18.685 $\pm$ 0.286 & nbo                & \uf \\
             &              &             & 19.024 $\pm$ 0.312 & 18.645 $\pm$ 0.278 & nbo                & \uf \\
 \\ 
 6C**1045+4459 & 10 48 32.4 & +44 44 28.0 & 18.581 $\pm$ 0.140 & 18.438 $\pm$ 0.202 & nbo                & \ui \\
 6C**1048+4434 & 10 51 26.52 & +44 18 21.2 & 18.805 $\pm$ 0.147 & 18.628 $\pm$ 0.210 & nbo                 & \ui \\
 6C**1050+5440 & 10 53 36.29 & +54 24 42.3 & 20.605 $\pm$ 0.224 & 20.164 $\pm$ 0.216 & 19.715 $\pm$ 0.264  & \uf \\
 6C**1052+4349$^{\dag}$ & 10 55 37.35 & +43 33 37.0 & 17.238 $\pm$ 0.026 & 17.142 $\pm$ 0.042 & 17.081 $\pm$ 0.074  & \uf \\
 6C**1056+5730$^{\dag}$ & 10 59 14.91 & +57 14 47.3 & 17.701 $\pm$ 0.041 & 17.562 $\pm$ 0.067 & 17.295 $\pm$ 0.101  & \uf \\
 \\
 6C**1100+4417 & 11 03 33.54 & +44 01 26.2 & 18.656 $\pm$ 0.100 & 18.334 $\pm$ 0.133 & 18.095 $\pm$ 0.196 & \uf \\
 6C**1102+4329 & 11 05 43.12 & +43 13 24.7 & 19.794 $\pm$ 0.740 & 19.793 $\pm$ 0.950 & 19.661 $\pm$ 1.297 & \uf \\
 6C**1103+5352 & 11 06 14.93 & +53 36 00.5 & 20.379 $\pm$ 0.374 & 20.142 $\pm$ 0.388 & nbo                & \uf \\
 6C**1105+4454 & 11 08 45.96 & +44 38 17.7 & 18.153 $\pm$ 0.048 & 18.086 $\pm$ 0.087 & 17.729 $\pm$ 0.127 & \uf \\ 
 6C**1106+5301 & 11 09 49.10 & +52 45 18.4 & 17.996 $\pm$ 0.071 & 17.609 $\pm$ 0.088 & 17.354 $\pm$ 0.129 & \uf \\
 \\
 6C**1112+4133 & 11 15 09.82 & +41 17 02.4 & 18.779 $\pm$ 0.098 & 18.435 $\pm$ 0.128 & 18.044 $\pm$ 0.162 & \uf \\
 6C**1125+5548 & 11 28 26.82 & +55 33 07.1 & 19.636 $\pm$ 0.250 & 19.680 $\pm$ 0.695 & nbo                & \uf \\
 6C**1132+3209 & 11 35 26.69 & +31 53 32.7 & 14.505 $\pm$ 0.003 & nbo                & nbo                & \uf \\
 6C**1135+5122 & 11 38 27.77 & +51 05 56.2 & 18.767 $\pm$ 0.119 & 18.554 $\pm$ 0.172 & nbo                 & \uf \\
 \\
 6C**1138+3309 & 11 41 25.98 & +32 52 11.8 & 18.574 $\pm$ 0.092 & 18.421 $\pm$ 0.142 & 18.014 $\pm$ 0.182    & \uf \\
             &              &             & 18.779 $\pm$ 0.111 & 18.435 $\pm$ 0.150 & 18.145 $\pm$ 0.244    & \uf \\
 6C**1138+3803$^{\dag}$ & 11 41 30.24 & +37 46 53.0 & 17.351 $\pm$ 0.028 & nbo                & nbo                 & \uf \\
 6C**1149+3509 & 11 51 50.75 & +34 53 01.6 & 19.240 $\pm$ 0.170 & 18.800 $\pm$ 0.247 & 18.729 $\pm$ 0.585  & \uf \\
 \hline
 \end{tabular}
 \end{center}
 {\caption{\label{tab:mag_journalk2} {\em continued.} 
}}
 \end{table*}

\begin{figure*}
%  \begin{center}
%%     \begin{minipage}[l]{0.49\linewidth}
%%       \epsfxsize=2.8in\epsfbox{fig1.1.ps}   
%%     \end{minipage}  
%%     \begin{minipage}[tl]{0.49\linewidth}
%%       \epsfxsize=2.8in\epsfbox{fig1.2.eps}   
%%     \end{minipage}  
%%   \end{center}
%%   \begin{center}
%%    \begin{minipage}[cl]{0.49\linewidth}
%%       \epsfxsize=2.8in\epsfbox{fig1.3.eps}   
%%     \end{minipage}  
%%     \begin{minipage}[cr]{0.49\linewidth}
%%       \epsfxsize=2.8in\epsfbox{fig1.4.eps}   
%%     \end{minipage}  
%%   \end{center}
%  \vskip 0.05cm
\caption{(a) The \kband images (greyscale) of the 68 6C** sources, with radio contours overlaid. The \kband images have been smoothed with a $\sigma = 1$\,pixel Gaussian for presentation purposes, unless otherwise stated. The black/white contours represent the VLA radio emission (with frequencies as given in Table~\ref{tab:im_journalk1}, column 5) or, when these are not  available, the radio maps from the 1.4\,GHz FIRST Survey. Contours are spaced at intervals separated by factors of 2 starting at: $0.4\,{\rm mJy\,beam^{-1}}$ for 6C**0714+4616; $0.3
2\,{\rm mJy\,beam^{-1}}$  for 6C**0717+5121; $0.4\,{\rm mJy\,beam^{-1}}$ for 6C**0726+4938; and, $0.2\,{\rm mJy\,beam^{-1}}$ for 6C**0737+5618.}
\label{fig:kband_images1}
\end{figure*}

\addtocounter{figure}{-1}

\begin{figure*}
%%   \begin{center}
%%    \begin{minipage}[bl]{0.49\linewidth}
%%       \epsfxsize=2.8in\epsfbox{fig1.5.eps}   
%%     \end{minipage}\hfill
%%     \begin{minipage}[r]{0.49\linewidth}
%%       \epsfxsize=2.8in\epsfbox{fig1.6.eps}   
%%     \end{minipage}\hfill
%%   \end{center}
%%   \begin{center}
%%     \begin{minipage}[l]{0.49\linewidth}
%%       \epsfxsize=2.8in\epsfbox{fig1.7.eps}   
%%     \end{minipage}\hfill
%%     \begin{minipage}[tl]{0.49\linewidth}
%%       \epsfxsize=2.8in\epsfbox{fig1.8.eps}   
%%     \end{minipage}\hfill
%%   \end{center}
%%   \begin{center}
%%    \begin{minipage}[cl]{0.49\linewidth}
%%       \epsfxsize=2.8in\epsfbox{fig1.9.eps}   
%%     \end{minipage}\hfill
%%     \begin{minipage}[cr]{0.49\linewidth}
%%       \epsfxsize=2.8in\epsfbox{fig1.10.eps}   
%%     \end{minipage}\hfill
%%   \end{center}
\caption{(b) Contours are spaced at intervals separated by factors of 2 starting at: $0.2\,{\rm mJy\,beam^{-1}}$ for 6C**0744+3702; $0.4\,{\rm mJy\,beam^{-1}}$ for 6C**0746+5445; $0.2\,{\rm mJy\,beam^{-1}}$ for 6C**0754+4640; $0.2\,{\rm mJy\,beam^{-1}}$  for 6C**0754+5019;  $0.4\,{\rm mJy\,beam^{-1}}$ for 6C**0801+4903; and, $0.4\,{\rm mJy\,beam^{-1}}$ for 6C**0810+4605. The images of 6C**0744+3702 and 6C**0754+5019 have been smoothed with a $\sigma = 2$\,pixel Gaussian.
}
\label{fig:kband_images2}
\end{figure*}

\addtocounter{figure}{-1}

\begin{figure*}
%%   \begin{center}
%%    \begin{minipage}[bl]{0.49\linewidth}
%%       \epsfxsize=2.8in\epsfbox{fig1.11.eps}   
%%     \end{minipage}\hfill
%%     \begin{minipage}[r]{0.49\linewidth}
%%       \epsfxsize=2.8in\epsfbox{fig1.12.eps}   
%%     \end{minipage}\hfill
%%   \end{center}
%%   \begin{center}
%%     \begin{minipage}[l]{0.49\linewidth}
%%       \epsfxsize=2.8in\epsfbox{fig1.13.eps}   
%%     \end{minipage}\hfill
%%     \begin{minipage}[tl]{0.49\linewidth}
%%       \epsfxsize=2.8in\epsfbox{fig1.14.eps}   
%%     \end{minipage}\hfill
%%   \end{center}
%%   \begin{center}
%%    \begin{minipage}[cl]{0.49\linewidth}
%%       \epsfxsize=2.8in\epsfbox{fig1.15.eps}   
%%     \end{minipage}\hfill
%%     \begin{minipage}[cr]{0.49\linewidth}
%%       \epsfxsize=2.8in\epsfbox{fig1.16.eps}   
%%     \end{minipage}\hfill
%%   \end{center}
\caption{(c) Contours are spaced at intervals separated by factors of 2 starting at: $0.4\,{\rm mJy\,beam^{-1}}$ for 6C**0813+3725; $0.4\,{\rm mJy\,beam^{-1}}$ for 6C**0824+5344; $0.32\,{\rm mJy\,beam^{-1}}$ for 6C**0829+3902; $0.32\,{\rm mJy\,beam^{-1}}$  for 6C**0832+4420; $0.2\,{\rm mJy\,beam^{-1}}$ for 6C**0832+5443; and, $0.32\,{\rm mJy\,beam^{-1}}$ for 6C**0834+4129. The image of 6C**0832+5443 has been smoothed with a $\sigma = 2$\,pixel Gaussian.}
\label{fig:kband_images3}
\end{figure*}

\addtocounter{figure}{-1}

\begin{figure*}
%%   \begin{center}
%%    \begin{minipage}[bl]{0.49\linewidth}
%%       \epsfxsize=2.8in\epsfbox{fig1.17.eps}   
%%     \end{minipage}\hfill
%%     \begin{minipage}[r]{0.49\linewidth}
%%       \epsfxsize=2.8in\epsfbox{fig1.18.eps}   
%%     \end{minipage}\hfill
%%   \end{center}
%%   \begin{center}
%%     \begin{minipage}[l]{0.49\linewidth}
%%       \epsfxsize=2.8in\epsfbox{fig1.19.eps}   
%%     \end{minipage}\hfill
%%     \begin{minipage}[tl]{0.49\linewidth}
%%       \epsfxsize=2.8in\epsfbox{fig1.20.eps}   
%%     \end{minipage}\hfill
%%   \end{center}
%%   \begin{center}
%%    \begin{minipage}[cl]{0.49\linewidth}
%%       \epsfxsize=2.8in\epsfbox{fig1.21.eps}   
%%     \end{minipage}\hfill
%%     \begin{minipage}[cr]{0.49\linewidth}
%%       \epsfxsize=2.8in\epsfbox{fig1.22.eps}   
%%     \end{minipage}\hfill
%%   \end{center}
%%   \vskip 0.05cm
\caption{(d) Contours are spaced at intervals separated by factors of 2 starting at: $0.2\,{\rm mJy\,beam^{-1}}$ for 6C**0848+4803; $0.4\,{\rm mJy\,beam^{-1}}$ for 6C**0848+4927; $0.2\,{\rm mJy\,beam^{-1}}$ for 6C**0849+4658; $0.4\,{\rm mJy\,beam^{-1}}$  for 6C**0854+3500; $0.4\,{\rm mJy\,beam^{-1}}$ for 6C**0855+4428; and, $0.4\,{\rm mJy\,beam^{-1}}$ for 6C**0856+4313.}
\label{fig:kband_images4}
\end{figure*}

\addtocounter{figure}{-1}

\begin{figure*}
%%   \begin{center}
%%    \begin{minipage}[bl]{0.49\linewidth}
%%       \epsfxsize=2.8in\epsfbox{fig1.23.eps}   
%%     \end{minipage}\hfill
%%     \begin{minipage}[r]{0.49\linewidth}
%%       \epsfxsize=2.8in\epsfbox{fig1.24.eps}   
%%     \end{minipage}\hfill
%%   \end{center}
%%   \begin{center}
%%     \begin{minipage}[l]{0.49\linewidth}
%%       \epsfxsize=2.8in\epsfbox{fig1.25.eps}   
%%     \end{minipage}\hfill
%%     \begin{minipage}[tl]{0.49\linewidth}
%%       \epsfxsize=2.8in\epsfbox{fig1.26.eps}   
%%     \end{minipage}\hfill
%%   \end{center}
%%   \begin{center}
%%    \begin{minipage}[cl]{0.49\linewidth}
%%       \epsfxsize=2.8in\epsfbox{fig1.27.eps}   
%%     \end{minipage}\hfill
%%     \begin{minipage}[cr]{0.49\linewidth}
%%       \epsfxsize=2.8in\epsfbox{fig1.28.eps}   
%%     \end{minipage}\hfill
%%   \end{center}
%%   \vskip 0.05cm
\caption{(e) Contours are spaced at intervals separated by factors of 2 starting at: $0.32\,{\rm mJy\,beam^{-1}}$ for 6C**0902+3827; $0.32\,{\rm mJy\,beam^{-1}}$ for 6C**0903+4251; $0.8\,{\rm mJy\,beam^{-1}}$ for 6C**0909+4317; $1.28\,{\rm mJy\,beam^{-1}}$  for 6C**0912+3913; $0.32\,{\rm mJy\,beam^{-1}}$ for 6C**0920+5308; and $0.8\,{\rm mJy\,beam^{-1}}$ for 6C**0922+4216.}
\label{fig:kband_images5}
\end{figure*}

\addtocounter{figure}{-1}

\begin{figure*}
%%   \begin{center}
%%    \begin{minipage}[bl]{0.49\linewidth}
%%       \epsfxsize=2.8in\epsfbox{fig1.29.eps}   
%%     \end{minipage}\hfill
%%     \begin{minipage}[r]{0.49\linewidth}
%%       \epsfxsize=2.8in\epsfbox{fig1.30.eps}   
%%     \end{minipage}\hfill
%%   \end{center}
%%   \begin{center}
%%     \begin{minipage}[l]{0.49\linewidth}
%%       \epsfxsize=2.8in\epsfbox{fig1.31.eps}   
%%     \end{minipage}\hfill
%%     \begin{minipage}[tl]{0.49\linewidth}
%%       \epsfxsize=2.8in\epsfbox{fig1.32.eps}   
%%     \end{minipage}\hfill
%%   \end{center}
%%   \begin{center}
%%    \begin{minipage}[cl]{0.49\linewidth}
%%       \epsfxsize=2.8in\epsfbox{fig1.33.eps}   
%%     \end{minipage}\hfill
%%     \begin{minipage}[cr]{0.49\linewidth}
%%       \epsfxsize=2.8in\epsfbox{fig1.34.eps}   
%%     \end{minipage}\hfill
%%   \end{center}
%%   \vskip 0.05cm
\caption{(f) Contours are spaced at intervals separated by factors of 2 starting at: $0.32\,{\rm mJy\,beam^{-1}}$ for 6C**0924+4933; $0.64\,{\rm mJy\,beam^{-1}}$ for 6C**0925+4155; $0.4\,{\rm mJy\,beam^{-1}}$ for 6C**0928+4203; $0.2\,{\rm mJy\,beam^{-1}}$  for 6C**0928+5557; $0.4\,{\rm mJy\,beam^{-1}}$ for 6C**0930+4856; and, $0.4\,{\rm mJy\,beam^{-1}}$ for 6C**0935+4348.}
\label{fig:kband_images6}
\end{figure*}

\addtocounter{figure}{-1}

\begin{figure*}
%%   \begin{center}
%%    \begin{minipage}[bl]{0.49\linewidth}
%%       \epsfxsize=2.8in\epsfbox{fig1.35.eps}   
%%     \end{minipage}\hfill
%%     \begin{minipage}[r]{0.49\linewidth}
%%       \epsfxsize=2.8in\epsfbox{fig1.36.eps}   
%%     \end{minipage}\hfill
%%   \end{center}
%%   \begin{center}
%%     \begin{minipage}[l]{0.49\linewidth}
%%       \epsfxsize=2.8in\epsfbox{fig1.37.eps}   
%%     \end{minipage}\hfill
%%     \begin{minipage}[tl]{0.49\linewidth}
%%       \epsfxsize=2.8in\epsfbox{fig1.38.eps}   
%%     \end{minipage}\hfill
%%   \end{center}
%%   \begin{center}
%%    \begin{minipage}[cl]{0.49\linewidth}
%%       \epsfxsize=2.8in\epsfbox{fig1.39.eps}   
%%     \end{minipage}\hfill
%%     \begin{minipage}[cr]{0.49\linewidth}
%%       \epsfxsize=2.8in\epsfbox{fig1.40.eps}   
%%     \end{minipage}\hfill
%%   \end{center}
%  \vskip 0.05cm
\caption{(g) Contours are spaced at intervals separated by factors of 2 starting at: $0.4\,{\rm mJy\,beam^{-1}}$ for 6C**0935+5548; $0.32\,{\rm mJy\,beam^{-1}}$ for 6C**0938+3801; $0.4\,{\rm mJy\,beam^{-1}}$ for 6C**0943+4034; $0.4\,{\rm mJy\,beam^{-1}}$  for 6C**0944+3946; $0.8\,{\rm mJy\,beam^{-1}}$ for 6C**0956+5735; and, $0.2\,{\rm mJy\,beam^{-1}}$ for 6C**0957+3955. The image of 6C**0938+3801 has been smoothed with a $\sigma = 2$\,pixel Gaussian.}
\label{fig:kband_images7}
\end{figure*}

\addtocounter{figure}{-1}

\begin{figure*}
%%   \begin{center}
%%    \begin{minipage}[bl]{0.49\linewidth}
%%       \epsfxsize=2.8in\epsfbox{fig1.41.eps}   
%%     \end{minipage}\hfill
%%     \begin{minipage}[r]{0.49\linewidth}
%%       \epsfxsize=2.8in\epsfbox{fig1.42.eps}   
%%     \end{minipage}\hfill
%%   \end{center}
%%   \begin{center}
%%     \begin{minipage}[l]{0.49\linewidth}
%%       \epsfxsize=2.8in\epsfbox{fig1.43.eps}   
%%     \end{minipage}\hfill
%%     \begin{minipage}[tl]{0.49\linewidth}
%%       \epsfxsize=2.8in\epsfbox{fig1.44.eps}   
%%     \end{minipage}\hfill
%%   \end{center}
%%   \begin{center}
%%    \begin{minipage}[cl]{0.49\linewidth}
%%       \epsfxsize=2.8in\epsfbox{fig1.45.eps}   
%%     \end{minipage}\hfill
%%     \begin{minipage}[cr]{0.49\linewidth}
%%       \epsfxsize=2.8in\epsfbox{fig1.46.eps}   
%%     \end{minipage}\hfill
%%   \end{center}
%%   \vskip 0.05cm
\caption{(h) Contours are spaced at intervals separated by factors of 2 starting at: $0.8\,{\rm mJy\,beam^{-1}}$ for 6C**1003+4827; $0.32\,{\rm mJy\,beam^{-1}}$ for 6C**1004+4531; $0.4\,{\rm mJy\,beam^{-1}}$ for 6C**1006+4135; $1.28\,{\rm mJy\,beam^{-1}}$  for 6C**1009+4327; $0.64\,{\rm mJy\,beam^{-1}}$ for 6C**1015+5334; and, $0.32\,{\rm mJy\,beam^{-1}}$ for 6C**1017+3436. The image of 6C**1009+4327 has been smoothed with a $\sigma = 2$\,pixel Gaussian.}
\label{fig:kband_images8}
\end{figure*}

\addtocounter{figure}{-1}

\begin{figure*}
%%   \begin{center}
%%    \begin{minipage}[bl]{0.49\linewidth}
%%       \epsfxsize=2.8in\epsfbox{fig1.47.eps}   
%%     \end{minipage}\hfill
%%     \begin{minipage}[r]{0.49\linewidth}
%%       \epsfxsize=2.8in\epsfbox{fig1.48.eps}   
%%     \end{minipage}\hfill
%%   \end{center}
%%   \begin{center}
%%     \begin{minipage}[l]{0.49\linewidth}
%%       \epsfxsize=2.8in\epsfbox{fig1.49.eps}   
%%     \end{minipage}\hfill
%%     \begin{minipage}[tl]{0.49\linewidth}
%%       \epsfxsize=2.8in\epsfbox{fig1.50.eps}   
%%     \end{minipage}\hfill
%%   \end{center}
%%   \begin{center}
%%    \begin{minipage}[cl]{0.49\linewidth}
%%       \epsfxsize=2.8in\epsfbox{fig1.51.eps}   
%%     \end{minipage}\hfill
%%     \begin{minipage}[cr]{0.49\linewidth}
%%       \epsfxsize=2.8in\epsfbox{fig1.52.eps}   
%%     \end{minipage}\hfill
%%   \end{center}
\caption{(i) Contours are spaced at intervals separated by factors of 2 starting at: $0.32\,{\rm mJy\,beam^{-1}}$ for 6C**1018+4000; $0.32\,{\rm mJy\,beam^{-1}}$ for 6C**1035+4245; $0.64\,{\rm mJy\,beam^{-1}}$ for 6C**1036+4721, $1.6\,{\rm mJy\,beam^{-1}}$ for 6C**1043+3714; 
$1.28\,{\rm mJy\,beam^{-1}}$ for 6C**1044+4938; and, $0.64\,{\rm
mJy\,beam^{-1}}$ for 6C**1045+4459.}
\label{fig:kband_images9}
\end{figure*}

\addtocounter{figure}{-1}

\begin{figure*}
%%   \begin{center}
%%    \begin{minipage}[bl]{0.49\linewidth}
%%       \epsfxsize=2.8in\epsfbox{fig1.53.eps}   
%%     \end{minipage}\hfill
%%     \begin{minipage}[r]{0.49\linewidth}
%%       \epsfxsize=2.8in\epsfbox{fig1.54.eps}   
%%     \end{minipage}\hfill
%%   \end{center}
%%   \begin{center}
%%     \begin{minipage}[l]{0.49\linewidth}
%%       \epsfxsize=2.8in\epsfbox{fig1.55.eps}   
%%     \end{minipage}\hfill
%%     \begin{minipage}[tl]{0.49\linewidth}
%%       \epsfxsize=2.8in\epsfbox{fig1.56.eps}   
%%     \end{minipage}\hfill
%%   \end{center}
%%   \begin{center}
%%    \begin{minipage}[cl]{0.49\linewidth}
%%       \epsfxsize=2.8in\epsfbox{fig1.57.eps}   
%%     \end{minipage}\hfill
%%     \begin{minipage}[cr]{0.49\linewidth}
%%       \epsfxsize=2.8in\epsfbox{fig1.58.eps}   
%%     \end{minipage}\hfill
%%   \end{center}
%%   \vskip 0.05cm
\caption{(j) Contours are spaced at intervals separated by factors of 2 starting at: $0.8\,{\rm mJy\,beam^{-1}}$ for 6C**1048+4434; $0.8\,{\rm mJy\,beam^{-1}}$ for 6C**1050+5440; $0.4\,{\rm mJy\,beam^{-1}}$ for 6C**1052+4349; 
$0.8\,{\rm mJy\,beam^{-1}}$ for 6C**1056+5730; $0.64\,{\rm
mJy\,beam^{-1}}$ for 6C**1100+4417; and, $1.6\,{\rm mJy\,beam^{-1}}$
for 6C**1102+4329. The images of 6C**1050+5440 and 6C**1102+4329 have
been smoothed with a $\sigma = 2$\,pixel Gaussian.}
\label{fig:kband_images10}
\end{figure*}

\addtocounter{figure}{-1}

\begin{figure*}
%%   \begin{center}
%%    \begin{minipage}[bl]{0.49\linewidth}
%%       \epsfxsize=2.8in\epsfbox{fig1.59.eps}   
%%     \end{minipage}\hfill
%%     \begin{minipage}[r]{0.49\linewidth}
%%       \epsfxsize=2.8in\epsfbox{fig1.60.eps}   
%%     \end{minipage}\hfill
%%   \end{center}
%%   \begin{center}
%%     \begin{minipage}[l]{0.49\linewidth}
%%       \epsfxsize=2.8in\epsfbox{fig1.61.eps}   
%%     \end{minipage}\hfill
%%     \begin{minipage}[tl]{0.49\linewidth}
%%       \epsfxsize=2.8in\epsfbox{fig1.62.eps}   
%%     \end{minipage}\hfill
%%   \end{center}
%%   \begin{center}
%%    \begin{minipage}[cl]{0.49\linewidth}
%%       \epsfxsize=2.8in\epsfbox{fig1.63.eps}   
%%     \end{minipage}\hfill
%%     \begin{minipage}[cr]{0.49\linewidth}
%%       \epsfxsize=2.8in\epsfbox{fig1.64.eps}   
%%     \end{minipage}\hfill
%%   \end{center}
%%   \vskip 0.05cm
\caption{(k) Contours are spaced at intervals separated by factors of 2 starting at: $0.64\,{\rm mJy\,beam^{-1}}$ for 6C**1103+5352; $1.6\,{\rm mJy\,beam^{-1}}$ for 6C**1105+4454; $0.64\,{\rm mJy\,beam^{-1}}$  for 6C**1106+5301; $0.8\,{\rm mJy\,beam^{-1}}$ for 6C**1112+4133; $1.28\,{\rm mJy\,beam^{-1}}$ for 6C**1125+5548; and, $0.64\,{\rm mJy\,beam^{-1}}$ for 6C**1132+3209. The images of 6C**1103+5352 and 6C**1125+5548 have been smoothed with a $\sigma = 2$\,pixel Gaussian. The image of 6C**1132+3209 has not been smoothed.}
\label{fig:kband_images11}
\end{figure*}

 \addtocounter{figure}{-1}

\begin{figure*}
%%   \begin{center}
%%    \begin{minipage}[bl]{0.49\linewidth}
%%       \epsfxsize=2.8in\epsfbox{fig1.65.eps}   
%%     \end{minipage}\hfill
%%     \begin{minipage}[r]{0.49\linewidth}
%%       \epsfxsize=2.8in\epsfbox{fig1.66.eps}   
%%     \end{minipage}\hfill
%%   \end{center}
%%   \begin{center}
%%     \begin{minipage}[l]{0.49\linewidth}
%%       \epsfxsize=2.8in\epsfbox{fig1.67.eps}   
%%     \end{minipage}\hfill
%%     \begin{minipage}[tl]{0.49\linewidth}
%%       \epsfxsize=2.8in\epsfbox{fig1.68.eps}   
%%     \end{minipage}\hfill
%%   \end{center}
\caption{(l) Contours are spaced at intervals separated by factors of 2 starting at: $0.64\,{\rm mJy\,beam^{-1}}$ for 6C**1135+5122, $0.8\,{\rm mJy\,beam^{-1}}$ for 6C**1138+3309; $0.8\,{\rm mJy\,beam^{-1}}$  for 6C**1138+3803; and, $0.64\,{\rm mJy\,beam^{-1}}$ for 6C**1149+3509.}
\label{fig:kband_images12}
\end{figure*}

\subsection{Redshifts and Line Parameters}\label{sec:redshifts}

Sixteen of the sources which have been observed with the WHT show
definite and/or plausible emission lines in their spectra.  A Gaussian
fit was made to all the lines with high enough signal-to-noise to
determine the line centre, FWHM and total line flux.  Continua were
fitted to the data by eye. All measurements were made on the FWZI
aperture spectra and prior to the smoothing process described in
Section~\ref{sec:spectra-datareduction}.  The emission-line
measurements are listed in Table~\ref{tab:spectra}.

To determine redshifts and identify the lines we compared our spectra
with the composite radio galaxy spectrum of McCarthy (1993),
which is based on observations of galaxies with $0.1 \,\, \ltsim \,\,
z \,\,\ltsim \,\, 3$. Of the objects with identifiable features in
their spectra three are classified as quasars (6C**0714+4616,
6C**0928+4203 and 6C**1036+3714) on the basis of broad-lines in the
spectrum and/or unresolved \kband emission.

 The redshifts listed on Table~\ref{tab:spectra} are based on the
strongest and cleanest line (i.e. a line profile not affected by
absorption) in each spectrum. In the case of quasars the redshifts are
based on narrow lines if possible.  The redshifts of six objects
(6C**0726+4938, 6C**0746+5445, 6C**0832+5443, 6C**1009+4327,
6C**1045+4459 and 6C**1102+4329) are based on a single emission-line
identification.  In five of these cases we associate it with
Ly$\alpha$, due to its structure and strength.  Redshifts based on the
Ly$\alpha$ line alone may not be the most accurate because in some
cases, either (i) absorption blueward of the line occurs shifting the
measured line centre to the red; or (ii) the line profile is severely
disturbed by strong absorption making redshift determination difficult
(van Ojik et al. 1997; Jarvis et al. 2003).
If there are other high signal-to-noise lines
present in the spectrum, the Ly$\alpha$ line is not used to estimate
redshifts.

We have estimated the spatial extent of some of the lines with higher
signal-to-noise ratio, typically: Ly$\alpha$, CIV and in three cases
[OII].  These were estimated by evaluating the full-width
zero-intensity of a cross-cut through the emission line, deconvolved
from the seeing (as in Rawlings, Eales \& Lacy 2001; Jarvis et al. 2001b).
The values
are presented in Table~\ref{tab:spectra}.

Using the NASA Extragalactic
Database\footnote{http://nedwww.ipac.caltech.edu/} (NED) we have
searched for known spectroscopic redshifts for the members of the 6C**
sample. We have found spectroscopic redshifts in the literature for
seven of the 6C** sources, one of which we have also obtained in this
paper (6C**0714+4616; Section~\ref{sec:redshifts}). The results of our
search
are summarized on Table~\ref{tab:literature}.

\subsection{Sources without a redshift}\label{sec:non-redshift}

For eleven of the sources which have been observed spectroscopically
it is not possible to determine a redshift. Nine sources
(6C**0737+5618, 6C**0744+3702, 6C**0754+4640, 6C**0813+3725,
6C**0829+3902, 6C**0848+4927, 6C**0925+4155, 6C**0938+3801 and
6C**1050+5440) do not show any reliable continuum or emission lines in
their spectra; whilst, two other sources (6C**0717+5121 and
6C**0902+3827) show weak red continuum but no obvious emission or
absorption features in their spectra.

  Although in most of these cases spectroscopy was taken ``blind'' we
  are confident that, with one possible exception, the respective
  near-infrared counterpart for each source was encompassed by the
  slit. The exception here is 6C**0754+4640. Its spectrum was taken
  pointed at the centre of the radio source with the slit aligned
  along the radio axis. If the radio source is identified by source
  (a) in our \kband image (Fig.~\ref{fig:kband_images1}), then the slit
  would have certainly targeted it. However, if the true identification
  is source (b), then that would have not been the case. Deeper
  spectroscopy with the slit passing through both \kband components
  will be necessary to establish their nature and relationship to the
  radio source.

\section{Notes on Individual Sources}\label{sec:notes}

\subsection{Near-infrared imaging}\label{sec:notes-imaging}
In this section we present notes on all individual sources in the 6C** sample.
These refer to the images presented in Fig.~\ref{fig:kband_images1}.
A summary of the sources with unresolved \kband emission is given in
Table~\ref{tab:unresolved}. These are most likely to be quasars.

\vskip 0.5cm

{\bf 6C**0714+4616} ($z = 1.466$) 
Our $K$-band image shows a bright unresolved identification co-spatial with the
radio core, as is expected for a quasar. This object, a highly
polarized red quasar, is discussed in great detail in De Breuck et
al. (1998).

{\bf 6C**0717+5121} A bright \kband \id is co-spatial (within the
astrometric uncertainty) with the southern lobe of a small radio
source. There is another source $\approx$ 4.5 arcsec to the west of
our identification which contaminates the photometry in an 8-arcsec
diameter aperture.

{\bf 6C**0726+4938} ($z = 1.203$\,?) 
A faint, diffuse K-band \id is co-spatial 
with the midpoint of the radio lobes for this small radio source.

{\bf 6C**0737+5618}
This is a compact radio source. We have a total of 2700 seconds
integration with NIRI on Gemini on this source with no apparent ID
down to a limiting $3\sigma$ magnitude of $K = 21.8$~mag in a 3-arcsec
diameter aperture. This is consistent with the results of De Breuck et
al. (2002),
who have also imaged this source, both in $K$- and \rband with NIRC on Keck and Kast on the the Lick 3m, respectively. Their \rband magnitude lower limit is $R > 24$~mag in a 4-arcsec diameter aperture.

{\bf 6C**0744+3702} ($z = 2.992$) Our \kband image shows a faint \id
co-spatial (within the astrometric uncertainty) with a double-lobed
radio source.  The \kband emission is diffuse and shows
sub-structure. Our \kband magnitude is consistent with the one
reported by De Breuck et al. (2002).

{\bf 6C**0746+5445} ($z = 2.156$) Our \kband image shows a
two-component (a, b) faint \id associated with the double-lobed radio
source. The \kband emission appears to be highly aligned with this
radio source. At the redshift of this object the H$\alpha$ emission
line is redshifted into the $K$-band. Therefore, it is possible that
the extended emission is associated with the emission line gas. There
is also another source (c) $\approx$ 3 arcsec to the north-west of our
identification which may be contributing some flux to the \kband
magnitude. This could plausibly be a companion galaxy to the
system. It has a \kband magnitude of 19.0 in a 3-arcsec diameter
aperture.

{\bf 6C**0754+4640} There are three possible faint \kband
identifications for this radio source present in our NIRI image. One
(a) is associated with the northern lobe along the radio axis, another
one (b) lies $\approx$ 2.5 arcsec to the south-west of the centre of
the radio source, and yet another one (c) is associated (within
astrometric uncertainty) with the southern lobe.  The $K$-band
magnitudes of (a) and (b) are very similar.
Faint diffuse emission surrounding both objects and features
suggestive of tidal tails in object (b), one of which seeming to reach
towards object (c), are indicative that the system may be undergoing a
major merger. The \kband magnitudes of both object (a) and (b) are
given in Table~\ref{tab:mag_journalk1}. Given their similarity, the
distribution in \kband magnitude of the sample is not affected
significantly by choosing one counterpart over the other. We assume
for the remainder of this paper that source (a) is the ID, as it lies
along the radio axis. However, further long-slit spectroscopy, with the
slit running through both objects, is required to confirm this.

{\bf 6C**0754+5019} ($z = 2.996$) The extremely faint \id (a) is
situated between the two radio lobes and it has been confirmed
spectroscopically (see notes on spectrum,
Section~\ref{sec:notes-spectra}). There is also another source (b)
$\approx$ 4 arcsec to the east of our \id which contaminates the
photometry in an 8-arcsec diameter aperture, and which could plausibly
be a companion galaxy.

{\bf 6C**0801+4903} There is a faint, diffuse \id co-spatial with the
radio source. Another slightly brighter source ($K = 19.1$~mag in a
5-arcsec aperture) $\approx$ 4.0 arcsec to the north-west west of our
identification, which could also be the ID, contaminates the
photometry in an 8-arcsec diameter aperture.

{\bf 6C**0810+4605} ($z = 0.620$) 
The \kband identification of this small double-lobed radio source is a
bright galaxy with an apparently disturbed morphology.

{\bf 6C**0813+3725} 
Our faint \kband \id is co-spatial (within the astrometric uncertainty) with a small radio source.

{\bf 6C**0824+5344} ($z = 2.824$)
This source has a double-lobed radio morphology, with a diffuse faint \kband
\id coincident with the radio core.

{\bf 6C**0829+3902} We find a faint \kband \id at the position of the
northern radio component.  There is also another source (b) $\approx$
4 arcsec to the south-west of our \id which contaminates the
photometry in an 8-arcsec diameter aperture.

{\bf 6C**0832+4420} 
The faint \kband \id for this source lies between the two radio lobes, along the radio axis but closer to the north-eastern one. 

{\bf 6C**0832+5443} ($z = 3.341$)
The faint \kband \id for this source lies between the two radio lobes but closer to the western lobe. There is another bright source $\approx$ 4 arcsec to the north of our \id which contaminates the photometry in an 8-arcsec diameter aperture. 

{\bf 6C**0834+4129} ($z = 2.442$) We find a faint \kband \id
co-spatial (within the astrometric uncertainty) with a small radio
source.

{\bf 6C**0848+4803} 
A bright \kband \id is co-spatial (within the astrometric uncertainty) with the radio core of a small double-lobed radio source.

{\bf 6C**0848+4927} 
A compact radio source with a faint \kband identification at its centre.

{\bf 6C**0849+4658}
Our \kband image shows a bright unresolved \id co-spatial with a small double radio source. It is possibly a quasar.

{\bf 6C**0854+3500} ($z = 2.382$)
We find a faint \kband \id co-spatial (within the astrometric uncertainty) with a small radio source.

{\bf 6C**0855+4428}
The \kband identification of this source lies at the radio core. There is a close  brighter source ($K = 17.1$ in an 5-arcsec aperture) $\approx$ 4 arcsec to the south-west, which is probably a foreground object and contaminates the photometry in an 8-arcsec diameter aperture.

{\bf 6C**0856+4313} ($z = 1.761$)
We find a faint \kband \id co-spatial (within astrometric uncertainty) with a small double radio source. The \kband emission appears to consist of a few knots of emission.

{\bf 6C**0902+3827}
We find a faint \kband identification at the centre of this double radio source. There is a brighter object $\approx$ 5 arcsec to the south of our ID, which contaminates the photometry in an 8-arcsec diameter aperture.

{\bf 6C**0903+4251} ($z = 0.907$)
Our \kband image shows a bright identification co-spatial with the
radio core. A knot of emission (b) is apparent in the image and it is
most likely associated with the system, e.g. a tidal tail. Fuzzy
emission around the source and hints of a disturbed morphology suggest
that it is the result of, or is currently undergoing, a merger event.

{\bf 6C**0909+4317}
The faint \kband \id (a) is situated  between the two radio lobes towards the north-eastern lobe. Towards the north-east of the \id there is a faint diffuse component (b), which is most likely part of the system and could possibly be a remnant of a merger, or a satellite object.

{\bf 6C**0912+3913}
Our image shows a faint \kband \id lying in between the radio lobes of this double radio source. The \id is close to a much brighter object, making magnitude estimation difficult.

{\bf 6C**0920+5308}
This source has a double-lobed radio morphology co-spatial with a bright resolved \kband ID. Apparent to the south-west of the host-galaxy is a knotty string of emission which is most likely a tidal tail, remnant of a major merger. Faint diffuse emission around the source is also consistent with this picture.

{\bf 6C**0922+4216} ($z = 1.750$)
The near-infrared \id for this source is the bright resolved object
$\approx 5$ arcsec to the north-west of the radio centroid. This
source has been identified as a quasar at $z = 1.750$ by Vigotti et
al. (1990).

{\bf 6C**0924+4933}
The radio source is co-spatial with a bright, resolved \kband identification.

{\bf 6C**0925+4155}
Our NIRC image shows a faint near-infrared \id co-spatial with the centre of this marginally resolved radio source.

{\bf 6C**0928+4203} ($z = 1.664$) The \kband \id for this source is
coincident with the southern lobe. 

{\bf 6C**0928+5557}
Our \kband image shows a bright \id co-spatial with the radio
source. Another \kband object $\approx$ 5 arcsec to the south-east of
our ID, with very similar \kband magnitude is plausibly a true
companion galaxy. There is faint diffuse emission surrounding the two
objects and the radio jet seems to reach towards the south-eastern
object. These two sources may well be in the process of merging,
although spectroscopy would be required to confirm such
hypothesis. Another possibility is that the south-eastern object is
the result of jet-induced star formation (e.g. Bicknell 2002).

{\bf 6C**0930+4856}
This small double source reveals a sub-structured \kband identification, with two peaks of emission.

\begin{figure}
%% \begin{center} 
%% {\hbox to 0.4\textwidth{ \epsfxsize=0.45\textwidth \epsfbox{fig2.eps}}}
\caption{A cut-out of the Gemini--NIRI $K$-band image of
  6C**0935+4348. The horizontal lines mark the width (to scale) and
  orientation of the 1.53-arcsec slit used in the optical
  spectroscopic observations of this source. The pointing position of
  spectroscopy and the approximate position of the spectrum along the
  slit are also marked. The distance between the pointing position and
  that of the spectrum is about 3.5 arcsec (from the astrometry on the
  2D spectrum). The
  distance between the source marked with 
 an A and the centre of the slit is 1.79 arcsec. 
}
 \label{fig:0935-scheme}
%\end{center}
\end{figure}

{\bf 6C**0935+4348} ($z = 2.321$?) This source has a slightly
extended\footnote{Deconvolved major axis of 5.1~arcsec, with a P.A. of
$\sim$~90$^{\circ}$ in the FIRST catalogue
(http://sundog.stsci.edu/cgi-bin/searchfirst).}  radio morphology,
centred at $\alpha =$ 09:38:21.41, $\delta = +$43:34:37.3. We have a
total of 2280 seconds integration with NIRI on Gemini on this source,
with no apparent \id around that position down to a limiting 3$\sigma$
magnitude of $K = 21.7$ in a 3-arcsec diameter aperture. Astrometry
from our two-dimensional spectrum places Ly$\alpha$ emission at a
position $\sim$ 3.5 arcsec to the west of the centre of the radio
source, along a slit (with a width of 1.53 arcsec) which was centred
at the radio source, with a position angle of 90$^{\circ}$
(see Fig.~\ref{fig:0935-scheme}). 
The infrared object $\approx$ 5.8 arcsec to the north-west (at $\alpha
=$ 09:38:20.99, $\delta = +$43:34:39.2) cannot be ruled out as the ID
because, although the slit does not pass through that object (marked
with an A in Fig.~\ref{fig:0935-scheme}), the Ly$\alpha$ emission can
be more extended than the \kband emission (e.g. Kurk et al. 2002).

{\bf 6C**0935+5548} We find a faint \kband \id at the position of the
eastern radio lobe. 

{\bf 6C**0938+3801} 
Our \kband image shows an object with a high-degree of sub-structure lying between the two radio lobes, but closer to the north-western one. 

{\bf 6C**0943+4034}
We find a faint \kband \id coincident with the centre of the radio source.

{\bf 6C**0944+3946}
We find a faint \kband \id at the centre of the radio source.

{\bf 6C**0956+4735} ($z = 1.026$)
Our \kband image shows a bright identification co-spatial with the radio core.

{\bf 6C**0957+3955}
The \kband \id for this source lies between the two radio lobes.

{\bf 6C**1003+4827} We find a bright, unresolved \kband identification
associated with the radio core, suggestive of a quasar.  The radio
structure of this source is also consistent with that of a quasar, in
that it shows a bright radio core and a one-sided jet.

{\bf 6C**1004+4531} We find a bright \kband \id coincident with the
position of the radio core. Towards the eastern lobe there is a
fainter clump of emission that is most probably associated with the
host galaxy, e.g. a region of enhanced star formation induced by the
jet (e.g. Bicknell 2002). This region is contributing to the \kband
flux.

{\bf 6C**1006+4135} We find a faint \kband \id at the position of the
north-western radio lobe. 

{\bf 6C**1009+4327} ($z = 1.956$) The extremely faint \id (a) is
 situated between the two radio lobes and it has been confirmed
 spectroscopically. There is also another source (b) $\approx$ 3
 arcsec to the north of our identification, which contaminates the
 photometry in apertures of 5-arcsec diameter and larger. Source (b)
 has similar \kband magnitude ($K = 20.44$ in 3-arcsec aperture) and
 could plausibly be a true companion galaxy.

{\bf 6C**1015+5334} A faint diffuse \kband \id is co-spatial with the
southern lobe of a double-lobed radio source.

{\bf 6C**1017+3436}
The \kband \id for this source lies between the two radio lobes.

{\bf 6C**1018+4000}
The \kband \id for this source is coincident with the radio core.

{\bf 6C**1035+4245} This small double-lobed radio source is co-spatial
with a bright resolved \kband \id.

{\bf 6C**1036+4721} ($z = 1.758$) A bright unresolved \kband \id lying
at the centre of the radio source, as is expected for a quasar.

{\bf 6C**1043+3714} ($z = 0.789$) Our \kband image shows a bright \id
coincident with the centre of the radio source. There is also a
fainter source $\approx 2$ arcsec to the south-west, which could be
contributing some flux to the \kband magnitude. The faint diffuse
emission surrounding the two objects suggests that they could be in
the process of merging. The optical spectrum of Allington-Smith et
al. (1985) shows an asymmetric extended region of
\mbox{[O II]~$\lambda$~3727\AA}\,\, emission, which the authors interpreted as
suggestive of interaction with a nearby companion (although no
companion was visible in their optical images). Our \kband image is
consistent with this scenario.

{\bf 6C**1044+4938} The faint \kband \id (a) is co-spatial with a
small double radio source. There are also two other faint sources
close to the identification. Both object (b) $\approx$ 4 arcsec to the
south-west, and (c) $\approx$ 5 arcsec to the north-west of our
identification may contribute some flux to the \kband magnitude in
apertures $> 5$ arcsec. Source (b) has similar \kband magnitude to
source (a), source (c) is much fainter.

{\bf 6C**1045+4459} ($z = 2.571$) The faint \kband \id (a) lies
between the radio lobes, closer to the south-eastern one. There is
also another source (b) $\approx$ 5 arcsec to the north-west of our
identification which may contribute some flux to the \kband
magnitude. Although the \kband magnitudes of both sources are very
similar, our spectroscopy shows that source (b) is a foreground galaxy
at $z = 0.883$ (if the line we detect is [O II]; see notes on spectrum
of this source, Section~\ref{sec:notes-spectra}).

{\bf 6C**1048+4434} Our \kband image shows three bright sources
associated with the radio source. We take the source lying in between
the radio lobes at the centre and along the radio axis to be our
plausible identification. The other two sources, lying to south-west
of the radio centroid, especially its nearest neighbour, cannot be
ruled out as the host-galaxy. All three sources have similar \kband
magnitudes.

{\bf 6C**1050+5440} Our \kband image shows a very faint \id coincident
with the centre of the radio source.

{\bf 6C**1052+4349} We find a bright, unresolved \kband \id co-spatial
with the centre of the radio source. This is possibly a quasar.

{\bf 6C**1056+5730}
This is a double-lobed radio structure with a bright unresolved \kband
\id close to its centre. This is possibly a quasar.

{\bf 6C**1100+4417}
We find a faint \kband identification at the centre of the radio source. 

{\bf 6C**1102+4329} ($z = 2.734$) 
This source has an extremely faint \kband identification, which lies in 
a position which is consistent with the centre of the radio source.

{\bf 6C**1103+5352} This source has an extremely faint \kband
identification lying at the centre of the radio map. The \id is close
to a much brighter object, making the magnitude estimation difficult.

{\bf 6C**1105+4454}
We find a bright \kband \id lying towards the north-western radio lobe.

{\bf 6C**1106+5301} This source has a faint \kband \id lying at the
centre of a slightly elongated radio source.

{\bf 6C**1112+4133} We find a faint \kband identification at the
centre of the double-lobed radio structure.

{\bf 6C**1125+5548} The faint \kband \id (a) lies at the position of
the southern radio lobe. There is also another source (b) $\approx$ 5
arcsec to the north-west of our identification, which contributes the
photometry in an 8-arcsec diameter aperture. This source has a very
similar \kband magnitude and could also be associated with the radio
source. Spectroscopic observations will be needed to confirm this.

{\bf 6C**1132+3209} ($z = 0.231$)
This is the brightest \kband source in our sample. Its \kband \id
appears to be a complex system of interacting galaxies. It is not
clear from Fig.~\ref{fig:kband_images1} but there are three peaks of
emission along the radio axis. The Nasa Extragalactic Database
associates this source with the MACS J1135.4+3153 galaxy cluster 
(Edge et al. 2003).

{\bf 6C**1135+5122} We find a faint \kband identification at the
centre of the radio source. Another source $\approx$ 4.5 arcsec to the
north-west contaminates the photometry in an 8-arcsec diameter
aperture.

{\bf 6C**1138+3309} This source has a faint \kband \id lying near the
centre of an elongated radio source.

{\bf 6C**1138+3803} A bright, unresolved \kband \id lies towards the
west of the radio centre. This is possibly a quasar.

{\bf 6C**1149+3509}
This source has a faint \kband \id lying at the centre of the radio source.

\begin{table*}
\scriptsize
\begin{center}
\begin{tabular}{lclcclcccc}
\hline
\mc{1}{c}{Source} & \mc{1}{c}{z} & \mc{1}{c}{Line} &  \mc{1}{c}{
$\lambda_{\mathrm rest}$} & \mc{1}{c}{$\lambda_{\mathrm obs}$} &
\mc{1}{c}{FWHM} & \mc{1}{c}{Flux} & \mc{1}{c}{$\log_{10} L_{\rm line}$} & \mc{1}{c}{Extent} & \mc{1}{c}{Galaxy /} \\
\mc{1}{c}{} & & & \mc{1}{c}{(\AA)} &
\mc{1}{c}{(\AA)} & \mc{1}{c}{(km~s$^{-1}$)} &
\mc{1}{c}{(Wm$^{-2}$)} & \mc{1}{c}{(W)} & \mc{1}{c}{(arcsec / kpc)} & \mc{1}{c}{Quasar} \\
\hline
 6C**0714+4616 & 1.466  & CIV        & 1549 & 3820 $\pm$ 1 & 0 -- 1200
 & 8.3E-19 $\pm$ 13\% & 36.05 & $<$ 3.5 / 30 & Q\\
              &       & MgII     & 2799 & 6897 $\pm$ 1 & 1150 -- 1300 & 2.1E-19 $\pm$ 14\% & 35.45\\
\\
 6C**0726+4938 & 1.203? & [OII]?      & 3727 & 8209 $\pm$ 1 & 450 -- 750  & 1.7E-19 $\pm$ 13\%& 35.15 & 5 / 41 & G\\

\\
 6C**0746+5445\ddag & 2.156 & Ly$\alpha$ & 1216 & 3838 $\pm$ 1 &
 $\sim$ 1200 & 5.2E-19 $\pm$ 24\%&  & $<$ 2.5 / 21 & G \\
\\
 6C**0754+5019  & 2.996 & Ly$\alpha$ & 1216 & 4858 $\pm$ 1 & 1100 --
 1500 & 8.3E-20 $\pm$ 22\% & 35.81& $<$ 3 / 23 & G \\
              &       & HeII       & 1640 & 6557 $\pm$ 1 & 1000 -- 1300 & 1.0E-19 $\pm$ 20\% & 35.89\\
\\
 6C**0810+4605 & 0.620 & MgII       & 2799 & 4527 $\pm$ 1 & 1600 -- 2100 & 1.1E-18 $\pm$ 11\% & 35.25 & & G \\
              &       & [OII]      & 3727 & 6039 $\pm$ 1  & 1150 -- 1550 & 9.1E-18 $\pm$ 13\% & 36.17 &  6 / 41\\
              &       & [NeIII]    & 3869 & 6264 $\pm$ 1  & 900 -- 1300 & 6.9E-19 $\pm$ 20\% & 35.05\\
              &       & [NeIII]    & 3968 & 6436 $\pm$ 1  & 300 -- 1000 & 4.5E-19 $\pm$ 20\% & 34.86\\
              &       & H$\gamma$  & 4340 & 7034 $\pm$ 1  & 1000 -- 1400 & 6.1E-19 $\pm$ 18\% & 34.99\\
              &       & [OIII]     & 4959 & 8026 $\pm$ 1  & 250 -- 800 & 8.5E-19 $\pm$ 34\% & 35.14\\
              &       & [OIII]     & 5007 & 8110 $\pm$ 1  & 900 -- 1200 & 3.3E-18 $\pm$ 14\% & 35.73\\
\\
 6C**0824+5344 & 2.824 & Ly$\alpha$$\dagger$ & 1216 & 4650 $\pm$ 1 & 0 -- 1000 & 1.3E-18 $\pm$ 5\% & 36.94 & 8 / 63 & G\\
\\
 6C**0832+5443 & 3.341 & Ly$\alpha$ & 1216 & 5279 $\pm$ 1 & 950 --
 1400 & 4.8E-19 $\pm$ 11\% & 36.68 & $<$ 3 / 22 & G\\
\\
 6C**0834+4129 & 2.442 & Ly$\alpha$ & 1216 & 4185 $\pm$ 1 & 0 -- 1100 & 4.8E-19 $\pm$ 13\% & 36.35 & 5 / 41 &G\\
              &       & CIV        & 1549 & 5335 $\pm$ 1 & 1250 -- 1600 & 3.2E-19 $\pm$ 35\% & 36.18\\
\\
 6C**0854+3500\ddag & 2.382 & Ly$\alpha$ & 1216 & 4113 $\pm$ 1 & $\sim 2100$ & & & $< 2$ / 16 &G\\
              &       & CIII]      & 1909 & 6456 $\pm$ 1 & $\sim 1000$\\
\\
 6C**0856+4313 & 1.761 & Ly$\alpha$ & 1216 & 3358 $\pm$ 1 & 0 -- 1650
 & 1.3E-18 $\pm$ 24\% & 36.44 & $<$ 3 / 25 & G\\
              &       & CIV        & 1549 & 4279 $\pm$ 1 & 0 -- 900 & 9.0E-20 $\pm$ 22\% & 35.28 & \\
\\
 6C**0928+4203 & 1.664 & Ly$\alpha$ & 1216 & 3240 $\pm$ 1 & 2300 -- 2800 & 4.9E-18 $\pm$ 6\% & 36.95 & 4.5 / 38 &Q\\
              &       & CIV        & 1549 & 4128 $\pm$ 1 & 800 -- 1500 & 4.0E-19  $\pm$ 6\% & 35.87 & 4 / 34\\
              &       & HeII       & 1640 & 4368 $\pm$ 1 & 1150 -- 1650 & 2.3E-19 $\pm$ 22\% & 35.63\\
              &       & CIII]$\dagger$ & 1909 & 5070 $\pm$ 1 & 2000 -- 2200 & 3.7E-19 $\pm$ 12\% & 35.83\\
              &       & CII]       & 2326 & 6206 $\pm$ 1 & 2150 -- 2300  & 3.3E-19 $\pm$ 16\% & 35.78\\    
              &       & MgII       & 2799 & 7447 $\pm$ 8 & $\sim$ 5600 & 1.1E-18 $\pm$ 11\% & 36.31\\
\\
 6C**0935+4348 & 2.321? & Ly$\alpha$? & 1216 & 4021 $\pm$ 1 & $\sim$ 1600 & 1.2E-18 $\pm$ 18\% & 36.70 & 8 / 65 & G\\
              &       & NV?         & 1240 & 4118 $\pm$ 1 & $<$ 500 & 2.4E-19 $\pm$ 20\%& 36.00\\
\\
 6C**1009+4327 & 1.956 & Ly$\alpha$ & 1216 & 3595 $\pm$ 1 & 2000 & 1.6E-19 $\pm$ 17\% & 35.64 & 7.0 / 59 & G\\
\\
 6C**1036+3714\ddag & 1.758 & Ly$\alpha$ & 1216 & 3337 $\pm$ 1 & $>$ 1900 & 2.8E-18 $\pm$ 11\% & & $< 4$ / 34 & Q\\
              &       & SiIV+OIV]& 1402 & 3867 $\pm$ 1 & $>$ 800 & 6.5E-19 $\pm$ 14\% & & \\
              &       & MgII       & 2799 & &  & & \\
\\
 6C**1045+4459 & 2.571 & Ly$\alpha$ & 1216 & 4341 $\pm$ 1 & 1600 -- 1900 & 6.6E-19 $\pm$ 9\% & 36.55  & 10 / 80 & G\\
 
(foregr. obj.)  & 0.883?& [OII]?     & 3727 & 7017 $\pm$ 1 & 750 -- 1000 & 1.8E-19 $\pm$ 17\% & 34.84 & 5 / 40 & G\\
\\
 6C**1102+4329 & 2.734 & Ly$\alpha$ & 1216 & 4541 $\pm$ 1 & 1200  & 6.8E-19 $\pm$ 2\% & 36.62 & 6.5 / 51 &G\\
\hline
\end{tabular}
\end{center}
\small {\caption{\label{tab:spectra} Redshifts and emission line
properties for the sources in the 6C** radio sample. The `?' symbol
denotes an uncertain line identification.  The $\dagger$ symbol means
that the line is contaminated by a cosmic ray. The \ddag \,\, symbol
means that the source was observed under non-photometric conditions.
For many of the uncertain lines in sources with known redshifts the
line diagnostics are not presented because of the low
signal-to-noise. For sources with redshifts based on uncertain lines,
we present the emission line data for the emission lines which are
likely to be real. Errors on the line fluxes represent the 1$\sigma$
uncertainty expressed as a percentage of the best line-flux estimate,
for the strongest lines these are dominated by roughly equal
contributions from uncertainties in fixing the local continuum level,
and from the absolute flux calibration (including plausible slit
losses).  Line widths were estimated from the FWHM of the best-fitting
Gaussian to each line, the lower value of a range assumes the
line-emitting region fills the slit, and the higher value assumes that
it is broadened only by the seeing.
The spatial extent of the emission lines were estimated by evaluating
the full-width zero-intensity of a cross-cut through the emission
line, deconvolved from the seeing.
There are no emission line fluxes available for 6C*0854+3500 due to
the lack of a spectrophotometric standard for the spectrum
presented. We do not measure the line parameters for MgII in
6C**1036+3714 as the line is severely extinguished by telluric
absorption (see notes on this source). We none the less present
line-flux measurements for the sources with non-photometric data
(6C**0746+5445 and 6C**1036+4721) but caution that these
should only be used to evaluate relative fluxes.}}
\end{table*}

\subsection{Optical Spectroscopy}\label{sec:notes-spectra}

In this section we present notes for all sources for which a redshift
has been obtained. These refer to the spectra presented in
Fig.~\ref{fig:spectra1}.

\vskip 0.5cm

{\bf 6C**0714+4616} Optical spectroscopy of this source has been
reported previously by De Breuck et al. (2001).
Definite CIV$\lambda$1549\,\AA\, and MgII$\lambda$2799\,\AA\, at $z =
1.466$ in our spectrum are in good agreement with this previous
observation. The HeII$\lambda$1640\,\AA\, line has FWHM $< 10$\,\AA\,
and could well be spurious. Infrared spectroscopy and
spectropolarimetric observations
have shown that this is a highly polarized red quasar (De Breuck et
al. 1998). Based solely on our spectrum it would have been difficult
to classify this source as a quasar given that, although
MgII$\lambda$2799\,\AA\, has a hint of a broad base, the narrow
component has a width of only about 1150--1300 km~s$^{-1}$.
However, this source has a very bright ($K = 16.3$ in an 8-arcsec
diameter aperture) unresolved identification
(Fig.~\ref{fig:kband_images10}), as is expected for a quasar.

{\bf 6C**0726+4938} This source displays an almost featureless faint
continuum. However, a definite line is detected in the far-red at
8209\,\AA, and we take it to be [OII]$\lambda$3727\,\AA\, at $z =
1.203$. This redshift is consistent with the \kband magnitude for this
object ($K = 18.2$ in an 8-arcsec aperture), but the lack of any
confirming features renders its value insecure.

{\bf 6C**0746+5445} The spectrum of this source contains a single
double-peaked line, which we take to be Ly$\alpha$ at $z = 2.156$. The
double-peaked structure is most probably due to HI absorption
associated with the host galaxy. The projected linear size of the
radio emission is small ($D \approx 26$ kpc). This is consistent with
the results of van Ojik et al. (1997) and Wilman et al. (2004),
in which small ($< 50$ kpc) radio sources are more likely to show
strong associated HI absorption. The spectrum presented in
Fig.~\ref{fig:spectra} was taken under non-photometric conditions.
The observations made in December 1998 yielded a blank field from
which it was not possible to extract a spectrum.

{\bf 6C**0754+5019} A blind spectrum was taken pointed at the centre
of the radio source with the slit aligned along the radio axis.  The
position of the line along the slit is consistent with the position of
our near-infrared identification (source (a) in
Fig.~\ref{fig:kband_images10}). We detect a weak-emission line
spectrum where the most prominent features are Ly$\alpha$ and
HeII$\lambda$1640\,\AA\, at $z = 2.996$. Somewhat unusual is that we
measure very similar fluxes for both of the lines. Comparison of the
Ly$\alpha$ and HeII line fluxes for radio galaxies using the ratios of
McCarthy (1993)
gives a Ly$\alpha$ to HeII ratio of 10:1. The strength of Ly$\alpha$
relative to HeII in our spectrum could be indicative of strong
Ly$\alpha$ attenuation, possibly by dust or HI. 
Objects with anomalously low Ly$\alpha$ emission
have been observed before. Dey, Spinrad \& Dickinson (1995)
reported a radio galaxy (MG 1019 + 0535) at $z = 2.76$ showing a
similarly weak Ly$\alpha$ to HeII ratio, which has been explained by
strong attenuation of Ly$\alpha$ by dust. However more relevant than
the Ly$\alpha$ line ratios, which are sensitive to neutral hydrogen
absorption, could be the NV/HeII and NV/CIV line ratios, which are
indicative of the nitrogen abundance (Villar-Martin et al. 1999; De
Breuck et al. 2000).
According to the interpretation of the NV/HeII
vs. NV/CIV diagram of Fosbury et al. (1999),
the nitrogen abundance could
be indicative of the level of chemical enrichment of the interstellar
medium produced by a stellar population composed of massive stars. Our
spectrum shows a hint of the NV line but it could also be a spurious
feature.  A deeper and higher resolution spectrum will be needed to
confirm the reality of this line, and constrain the NV/CIV and NV/HeII
ratios. Our object is also one of the faintest sources in the sample,
with $K = 20.6$ in a 5-arcsec diameter aperture, which is unusually
faint for this redshift. Again, this is suggestive of a dusty screen.
Sub-millimetre photometry would be helpful in establishing the
nature of this source.

%########################################
\begin{figure*}
%%   \begin{center}
%%     \begin{minipage}[r]{0.50\linewidth}
%%       \psfig{figure=fig3.1.eps,width=0.55\textwidth,angle=90}
%%     \end{minipage}\hfill
%%     \begin{minipage}[r]{0.50\linewidth}
%%       \psfig{figure=fig3.2.eps,width=0.55\textwidth,angle=90}
%%     \end{minipage}\hfill
%%   \end{center}
%%   \begin{center}
%%     \begin{minipage}[r]{0.50\linewidth}
%%       \psfig{figure=fig3.3.eps,width=0.55\textwidth,angle=90}
%%     \end{minipage}\hfill
%%     \begin{minipage}[r]{0.50\linewidth}
%%       \psfig{figure=fig3.4.eps,width=0.55\textwidth,angle=90}
%%     \end{minipage}\hfill
%%   \end{center}
%%   \begin{center}
%%     \begin{minipage}[r]{0.50\linewidth}
%%       \psfig{figure=fig3.5.eps,width=0.55\textwidth,angle=90}
%%     \end{minipage}\hfill
%%     \begin{minipage}[r]{0.50\linewidth}
%%       \psfig{figure=fig3.6.eps,width=0.55\textwidth,angle=90}
%%     \end{minipage}\hfill
%%   \end{center}
%%   \begin{center}
%%     \begin{minipage}[r]{0.50\linewidth}
%%       \psfig{figure=fig3.7.eps,width=0.55\textwidth,angle=90}
%%     \end{minipage}\hfill
%%     \begin{minipage}[r]{0.50\linewidth}
%%       \psfig{figure=fig3.8.eps,width=0.55\textwidth,angle=90}
%%     \end{minipage}\hfill
%%   \end{center}
\caption{(a) Spectra of the 6C** WHT--ISIS targets with definite or possible spectral features. Uncertain (possible) emission lines are marked with a `?'. The synthetic aperture used to extract 1D spectra from the 2D data was defined by the full-width at zero-intensity of a cross-cut through the data, excepting the cases of 6C**0726+4938 and 6C**0754+5019 for which a full-width at half-maximum aperture was used. The shadowed region in the spectrum of 6C**0810+4605 shows the wavelengths affected by severe fringing in the CCD. The flux density scale for the spectrum of 6C**0854+3500 is measured in W m$^{-2}$ \AA$^{-1}$\, with arbitrary normalisation, due to the lack of a spectrophotometric standard. The observations of 6C**0746+5445 and 6C**1036+4721 were made under non-photometric conditions.}
\label{fig:spectra1}
\end{figure*}

\addtocounter{figure}{-1}

\begin{figure*}
%%   \begin{center}
%%     \begin{minipage}[l]{0.50\linewidth}
%%       \psfig{figure=fig3.9.eps,width=0.55\textwidth,angle=90}
%%     \end{minipage}\hfill
%%     \begin{minipage}[tl]{0.50\linewidth}
%%       \psfig{figure=fig3.10.eps,width=0.55\textwidth,angle=90}
%%     \end{minipage}\hfill
%%   \end{center}
%%   \begin{center}
%%     \begin{minipage}[l]{0.50\linewidth} 
%%       \psfig{figure=fig3.11.eps,width=0.55\textwidth,angle=90}
%%     \end{minipage}\hfill
%%     \begin{minipage}[tl]{0.50\linewidth}
%%     \psfig{figure=fig3.12.eps,width=0.55\textwidth,angle=90}
%%     \end{minipage}\hfill
%%   \end{center}
%%   \begin{center}
%%     \begin{minipage}[l]{0.50\linewidth}
%%       \psfig{figure=fig3.13.eps,width=0.55\textwidth,angle=90}
%%     \end{minipage}\hfill
%%     \begin{minipage}[tl]{0.50\linewidth}
%%       \psfig{figure=fig3.14.eps,width=0.55\textwidth,angle=90} 
%%     \end{minipage}\hfill
%%   \end{center}
%%     \begin{minipage}[l]{0.50\linewidth}
%%      \psfig{figure=fig3.15.eps,width=0.55\textwidth,angle=90}
%%     \end{minipage}\hfill
%%     \begin{minipage}[l]{0.50\linewidth}
%%       \psfig{figure=fig3.16.eps,width=0.55\textwidth,angle=90}
%%     \end{minipage}\hfill
\caption{(b){\em continued}}
\label{fig:spectra}
\end{figure*}

\addtocounter{figure}{-1}

\begin{figure*}
%{\hbox to 1.0\textwidth{\psfig{figure=fig3.17.eps,width=0.275\textwidth,angle=90} }}
\caption{(c){\em continued}}
\label{fig:spectra}
\end{figure*}
%#######################################################

{\bf 6C**0810+4605} We detect strong [OII]$\lambda$3727\,\AA\, at $z = 0.620$. Several other lines: MgII$\lambda$2799\,\AA,
[NeIII]$\lambda$3869\,\AA, [NeIII]$\lambda$3968\,\AA, H$\gamma$,
[0III]$\lambda$4959\,\AA, [0III]$\lambda$5007\,\AA \, and possible
H$\beta$, confirm this redshift. This source shows a low level of
ionization, as indicated by a [OII]/[OIII] ratio of about 2.8. This,
together with the indication from our \kband image that this source
could be undergoing a major merger, suggests that shock heating may be
the dominant excitation mechanism operating in this source. The close
alignment between the optical structure and the radio source is also
consistent with this picture (Tadhunter 2002).
It is possible that
jet-induced shocks are driving a starburst over the whole galaxy.

{\bf 6C**0824+5344} 
We find a strong line at 4650\,\AA, which we take to be Ly$\alpha$ at
$z = 2.824$. Probable HeII$\lambda$1640\,\AA\, and
CIII]$\lambda$1909\,\AA, and the \kband magnitude ($K = 19.4$ in a
  8-arcsec diameter aperture) are consistent with this redshift. The
  single blue arm exposure containing the Ly$\alpha$ emission has a
  cosmic ray in close proximity to the line, which may contribute to
  the flux measurement in Table~\ref{tab:spectra}.

{\bf 6C**0832+5443} There is just one emission feature and no
continuum in the spectrum of this object. The emission line shows a
double-peaked structure.  From its wavelength, the lack of continuum
and structure we take it to be Ly$\alpha$ at $z = 3.341$. Again, the
structure of the line is probably due to HI absorption in the host
galaxy, and the size of the radio source ($D \approx 48$ kpc) is within
the 50 kpc limit of van Ojik et al. (1997).

{\bf 6C**0834+4129} The spectrum shows a clear emission line at
4185\,\AA, which we identify with Ly$\alpha$ at $z = 2.442$. Further
faint lines corresponding to CIV$\lambda$1549\,\AA\, and probable
HeII$\lambda$1640\,\AA\, confirm this redshift.

\begin{table}
\scriptsize
\begin{center}
\begin{tabular}{lclcl}
\hline
\mc{1}{c}{Name} & \mc{1}{c}{z} & \mc{1}{c}{Other}&\mc{1}{c}{Gal./} &\mc{1}{c}{Ref.}\\
\mc{1}{c}{} & \mc{1}{c}{} & \mc{1}{c}{Name}&\mc{1}{c}{Quasar} &\mc{1}{c}{}\\
\hline
6C**0714+4616 & 1.462 & WN J0717+4611 & Q &  1\\
6C**0744+3702 & 2.992 & WN J0747+3654 & G &  1\\
6C**0903+4251 & 0.907 & B3 0903+428   & G &  2\\
6C**0922+4126 & 1.750 & B3 0922+422   & Q &  3\\
6C**0956+4735 & 1.026 & 4C +47.31     & G &  2\\
6C**1043+3714 & 0.789 & 4C +37.28     & G &  4\\
6C**1132+3209 & 0.231 & 2MASX J11352669+3153324 & G & 5\\
\hline 
\end{tabular}
\end{center}
{\caption{\label{tab:literature} Spectroscopic redshifts from the
    literature. REFERENCES: 1. De Breuck et al. (2001); 2. McCarthy (1990); 3.
    Vigotti et al. (1990); 4. Allington-Smith et al. (1985);
    5. Brinkmann et al. (2000).}}
\end{table}

{\bf 6C**0854+3500} 
The presence of Ly$\alpha$ and
CIII]$\lambda$1909\,\AA \, in our spectrum confirm this as a radio
galaxy at $z = 2.382$. Again, we detect a double-peaked line structure
associated with a very small radio source ($D \simeq 9$~kpc).  The
lack of a spectrophotometric standard for the set-up used to take this
spectrum makes it impossible to flux calibrate it. Hence, the
flux-density scale presented in Fig.~\ref{fig:spectra} can only be
used to estimate relative fluxes.

{\bf 6C**0856+4313} 
Strong Ly$\alpha$, along with weak CIV$\lambda$1549\,\AA, possible NV$\lambda$1240\,\AA\, and CIII]$\lambda$1909\,\AA\, confirm this as a radio galaxy at $z = 1.761$.

{\bf 6C**0928+4203} This source has a rich emission-line spectrum. Six
definite emission lines place it at $z = 1.664$. The spectrum shows
broad MgII$\lambda$2799\,\AA\, ($\sim$ 5600 km~s$^{-1}$), but our \kband image
(Fig.~\ref{fig:kband_images10}) shows a faint ($K = 18.4$ in an
8-arcsec diameter aperture) resolved source, hence we classify this
source as reddened quasar. The redshift we obtain is consistent with
the \kband magnitude and well within the scatter of the $K-z$
relation for radio galaxies. The broad MgII emission could therefore
be a scattered quasar component.

{\bf 6C**0935+4348} We detect Ly$\alpha$ and NV$\lambda$1240\, \AA \,
in our high-resolution ($\sim$ 5.5 \AA) spectrum. Before we proceed to
discuss it, we caution that these line identifications are uncertain,
and the near-infared identification for this source is not secure (see
note on Section~\ref{sec:notes-imaging}). The Ly$\alpha$ to NV ratio
is particularly low in this source (5:1), when compared with the much
higher ratio (20:1) derived by McCarthy (1993) for radio
galaxies. Albeit this line ratio is suggestive of a quasar, the
observed width of the Ly$\alpha$ line (of about 1600 km~s$^{-1}$) is
characteristic of a radio galaxy. Similar cases of high NV to
Ly$\alpha$ ratios have been found before in radio galaxies at $z \sim
2-3$, such as TX 0211-122 (van Ojik et al. 1994) and TXS J2353-0002
(De Breuck et al. 2001).  TX 0211-122 has been explained as a dusty
galaxy, possibly undergoing a massive starburst based on the
similarity of its spectrum to that of the $z = 2.286$ ultraluminous
IRAS galaxy F10214+4724 (Rowan-Robinson et al. 1991).  It has been
argued the starburst would produce the dust responsible for
attenuating the Ly$\alpha$ emission and a relative overabundance of
nitrogen. Another important feature in the spectrum of our source is
the absence of CIV$\lambda$1549\,\AA, which we would expect to be
present at around 5144\, \AA. This is indicative of a high NV/CIV
ratio, which is a better indicator of nitrogen overabundance (see also
the notes for 6C**0754+5019).

{\bf 6C**1009+4327}
The only emission line present in the spectrum of this object is detected in the far-blue. We take it to be Ly$\alpha$ at $z = 1.956$. 
We detect continuum emission in both the red and blue arm exposures but no other identifiable features. 
The position of the line along the slit is consistent with the
position of our near-infrared counterpart (source (a) in 
Fig.~\ref{fig:kband_images10}), which is unusually faint ($K = 20.5$ in
a 3-arcsec diameter aperture) for this redshift.

{\bf 6C**1036+4721} This is a reddened quasar at \mbox{$z = 1.758$}. We find strong
Ly$\alpha$ in the far-blue of the spectrum, along with fainter SiIV +
OIV]$\lambda$1402\,\AA\, and  MgII$\lambda$2799\,\AA. The blue wing of
  the MgII emission line is severely extinguished by the 7600\,\AA\,
  A-band telluric absorption line system. 
  Telluric absorption
  is also found at $\sim$ 6821\,\AA. Because of the strong absorption
  it is not possible to derive accurate line parameters for the MgII
  line. Moreover it is not possible to estimate its width, although the
  line seems to have a broad base. We none the less classify this source
  as a quasar, given that its \kband identification is a bright ($K =
  17.0$ in an 8-arcsec diameter aperture) unresolved source. We note
  also that the optical spectroscopic observations of this source were made under non-photometric conditions.

{\bf 6C**1045+4459} 
A blind spectrum was taken pointed at the centre of the radio source
with the slit aligned along the radio axis. The spectrum shows no
continuum and only a single strong, spatially extended line which we
associate with Ly$\alpha$ at $z = 2.571$. The position of the line
along the slit is consistent with the position of our near-infrared
identification (source (a) in Fig.~\ref{fig:kband_images10}). 
We also detect the object $\approx$ 5 arcsec to north-west of the radio galaxy ($K = 18.73$ in a 5 arcsec aperture), which is visible in our \kband image (source (b) in Fig.~\ref{fig:kband_images10}).
A strong emission line at 7017 \AA, which we associate with [OII]$\lambda$3727\,\AA, and the blue continuum present in the spectrum of this object (also shown in Fig.~\ref{fig:spectra}) leads us to conclude that it is a foreground galaxy at $z = 0.883$, and is not associated with the radio emission.

{\bf 6C**1102+4329} 
Another object for which there is just one emission feature and no
continuum in the spectrum. From its wavelength and the lack of
continuum we take it to be Ly$\alpha$ at $z = 2.734$.

\section{Discussion}\label{sec:discussion}

\subsection{Number of chance coincidences expected among the near-infrared identifications}

We have identified all but two of the sources in the 6C** sample, down
to a limiting magnitude of $K \sim 21$~mag (8-arcsec diameter
aperture). Note, however, the caveat that one of the
non-detected sources (6C**0935+4348) has an uncertain identification
(see notes on this source in Section~\ref{sec:notes-imaging}).  Using
K-band number counts from Gardner, Cowie \& Wainscoat (1993) and
assuming an average search radius of 2.5 arcsec, we now estimate the
number of chance coincidences expected inside the matching area. This
is a function of K-band magnitude reached.

Deep infrared surveys (e.g. Gardner, Cowie \& Wainscoat 1993)
%\citep*[e.g.][]{GCW93} 
have shown that there are about $7.8 \times
10^{4}$ galaxies per square degree brighter than $K =
21$\,mag. Therefore, we expect 0.12 chance coincidences within a
2.5\,arcsec matching radius in the fields imaged with NIRI and NIRC.
With UFTI we reach $K \sim 19.5$\,mag (8-arcsec diameter aperture)
with the shortest exposures (540 sec.) and $K \sim 20$\,mag (8-arcsec
diameter aperture) with the deepest ones (3240 sec.). The $K$-band
number counts are down to $\sim 2.9 \times 10^{4}$ and $\sim 4.0
\times 10^{4}$ per square degree, respectively, and the chance
coincidences expected within the same matching radius are 0.06 and
0.04. Finally, with UIST we reach $K \sim 19$\,mag (8-arcsec diameter
aperture), to which the number counts are $\sim 2.0 \times 10^{4}$ per
square degree. Therefore, we expect 0.03 chance coincidences between a
radio and near-infrared source within a 2.5\,arcsec matching radius.

\subsection{\kband magnitude distribution}

\begin{figure}
\begin{center} 
%{\hbox to 0.4\textwidth{ \epsfxsize=0.45\textwidth \epsfbox{fig4.eps}}}
\caption{Histogram of the \kband magnitudes, measured in a 3-arcsec
  diameter aperture, of the 6C** sample (solid line) and the 6C*
  sample (dashed line). The dotted line represents the distribution of
  $K$-band magnitudes, measured in a 5-arcsec aperture, for the
  sources in the 7C-I and 7C-II regions of the 7C Redshift Survey. The
  bin width in each histogram is \mbox{$\Delta K = 0.5$~mag}. The
  non-detected sources (6C**0737+5618 and 6C**0935+4348) are
  represented by arrows.}
 \label{fig:histmag}
\end{center}
\end{figure}

In Fig.~\ref{fig:histmag} we compare the distribution of
$K$-magnitudes of the 6C** sample with those of the similarly selected
6C* sample (Jarvis et al. 2001a) and of the 7C-I and 7C-II regions of
7CRS (Willott et al. 1998, 2002, 2003). The later were selected at the
same flux level as 6C** but without any filtering.\footnote{The 7C-I
region covers an area of sky of 0.0061\,sr and contains 37 sources
with $S_{151} > 0.5$\,Jy. The 7C-II region cover an area of sky
0.0069\,sr and contains 38 sources with $S_{151} > 0.5$\,Jy.}.  The
median \kband magnitude of the 6C** sample is $\approx 18.7$~mag
(3-arcsec diameter aperture), with a maximum magnitude of $\gtrapprox
21.7$~mag, while the 6C* sample has a median \kband magnitude of
$\approx 18.8$~mag in the same aperture, with the faintest magnitude
being $\approx 20$~mag. It can be seen that although the distributions
of both samples peak at the same median magnitude, that of the 6C**
sample extends both to brighter and fainter magnitudes, with a
slightly asymmetrical tail extending towards bright magnitudes ($K
\simeq 14.5$~mag). A Kolmogorov-Smirnov test shows that the two
datasets are consistent with being drawn from the same underlying
distribution with probability $p = 0.51$.  Therefore, we expect the
6C** sample to have a similar redshift distribution to that of the 6C*
sample.  The 6C* sample has a median redshift of $z \simeq 1.9$, with
a minimum redshift of 0.513 and maximum redshift of 4.410 (Jarvis et
al. 2001b).  The median \kband magnitude of the combined 7-CI and
7C-II regions is $\approx 17.4$~mag (5-arcsec aperture). This shows
that the filtering criteria employed to select the 6C** sample is
being effective in biasing the sample towards fainter sources, and
possibly higher redshifts. A full derivation of redshift estimates for
the members of the 6C** sample, based on their \kband magnitudes, will
be presented in Paper II, along with a detailed analysis of their
space density.

\begin{table}
\small
\begin{center}
\begin{tabular}{lcl}
\hline
\mc{1}{c}{Name} & \mc{1}{c}{$K$} & \mc{1}{c}{Comments}\\
\mc{1}{c}{} & \mc{1}{c}{(mag)} & \mc{1}{c}{}\\
\hline
6C**0717+5121 & 17.79(5)   &  Faint red continuum, no lines\\ 
6C**0737+5618 & $>21.8$(3) &  Blank\\
6C**0744+3702 & 19.44(8)   &  Blank\\
6C**0754+4640 & 20.09(5)   &  Blank\\
6C**0813+3725 & 18.50(8)   &  Blank\\
6C**0829+3902 & 19.41(5)   &  Blank\\
6C**0848+4927 & 18.22(8)   &  Blank\\
6C**0902+3827 & 19.29(5)   &  Faint red continuum, no lines\\
6C**0925+4155 & 20.30(5)   &  Blank\\
6C**0938+3801 & 18.13(8)   &  Blank\\
6C**1050+5440 & 19.71(8)   &  Blank\\
\hline 
\end{tabular}
\end{center}
{\caption{\label{tab:no-redshift} Attempted spectroscopic observations not yielding a redshift.}}
\end{table}

\subsection{Sources without a redshift}\label{sec:non-redshift2}
We now discuss the sources for which we could not determine a redshift
from optical spectroscopy (listed in Table~\ref{tab:no-redshift}).  It
is likely that some of the these sources lie in the redshift range
$1.2 < z < 1.8$, where it is difficult to measure redshifts, due to
the absence of the strongest lines of the radio galaxy spectrum in the
optical wavelength region (e.g. Lacy et al. 1999b).
%The $K$-magnitudes of
%6C**0717+5121, 6C**0813+3725, 6C**0848+4927, 6C**0902+3827 and
%6C**0938+3801 are consistent with
%this explanation. 
Moreover, some may have SEDs which are similar to those of the $1 < z
< 2$ EROs, found in the 7CRS redshift survey (Willott et al. 1999),
which may be counterparts to radio galaxies which lack strong emission
lines in their spectra. 
Other sources will be at $z > 1.8$, but 
%The remaining sources, with $K \,\gtsim \,
%19.5$, are most likely at $z > 2$ and 
have a weak emission line
spectrum, which falls below our detection limit ($\sim 5 \times
10^{-20}$\,W\,m$^{-2}$ in $\sim$ 30~min. exposures).

\begin{table}
\small
\begin{center}
\begin{tabular}{lcc}
\hline
\mc{1}{c}{Name} & \mc{1}{c}{$K$} & \mc{1}{c}{$z$}\\
\mc{1}{c}{} & \mc{1}{c}{(mag)} & \mc{1}{c}{}\\
\hline
6C**0714+4616 & 16.3(8) & 1.466\\ 
6C**0849+4658 & 17.3(8) & -- \\
6C**0922+4216 & 15.9(8) & 1.750 \\
6C**1003+4827 & 16.9(8) & -- \\
6C**1036+4721 & 17.1(8) & 1.758 \\
6C**1052+4349 & 17.1(8) & -- \\
6C**1056+5730 & 17.3(8) & --\\
6C**1138+3803 & 17.3(3) & -- \\
\hline
\end{tabular}
\end{center}
{\caption{\label{tab:unresolved} Summary of unresolved \kband
    sources. We assume that these sources are quasars based on their 
    unresolved continuum emission. 
    Redshifts are given for sources where spectroscopy is
    available. We note that in all these cases spectroscopy has confirmed 
    the source 
    as a quasar. 
}}
\end{table}

 For all the sources with no measured redshifts it is possible that
 deeper spectroscopy will enable redshift determination. Indeed, this
 is the case of 6C**0744+3702, which is not detected in our one hour
 exposure spectrum (on WHT), but for which Keck spectroscopy (De
 Breuck et al. 2001)
 finds
 a weak-emission-line spectrum at $z = 2.992$ (see also
 Table~\ref{tab:literature}). However, as found by De Breuck et al. (2001)
some steep-spectrum radio galaxies are not detected even in long
 exposures ($\simeq 1-2$~hrs) with 8-10 metre class telescopes. These
 sources seem to be preferentially characterised by compact radio
 morphologies ($\theta \,\, \ltsim \,\, 2-3$~arcsec).

One of the objects we targeted (6C**0737+5618) has these
characteristics, and is also one of the faintest (\mbox{$K > 21.8$\,mag} in
a 3-arcsec aperture) objects in our sample. Several attempts in
different observing runs were made at obtaining optical spectroscopy
for this object. Different position angles and increasingly higher
exposure times were used each time. None the less this source remained
undetected in our WHT spectra.  Its \kband magnitude suggests a very
high redshift, which could well be above the optical detection limit
at $z \,\, \gtsim \,\, 6.5$, if the $K-z$ relation holds to such high
redshifts. Moreover, if a source is at $z \,\gtsim\, 6$, i.e. within
the epoch of reionization, it may have its Ly$\alpha$ emission
severely suppressed by the intervening intergalactic medium (IGM, Gunn
\& Peterson 1965, Becker et al. 2001). However, it is possible that
Ly$\alpha$ remains visible even when it is embedded in a neutral IGM
(e.g. Haiman 2002; Santos 2004; Furlanetto, Zaldarriaga \& Hernquist
2006).  Another possibility is that 6C**0737+5628 is at a lower
redshift but heavily obscured (e.g. like WN J0305+3525 in Reuland et
al. 2003), possibly due to it being observed shortly after the
jet-triggering event.

Given the nature of our selection criteria, i.e. a small radio angular
size, the galaxies which are at high-redshift ($z \gtrsim 2 $) in our
sample are preferentially young and seen shortly after the
jet-triggering event (Blundell \& Rawlings 1999).  The projected
linear size of the radio emission, which may be used as an age
indicator for the radio source, (Blundell, Rawlings \& Willott 1999)
is also found to anti-correlate with the 850\,$\mu$m flux-density
(Willott et al. 2002), which in turn is an indicator of the dust
content of a galaxy. We speculate that the spectra for some of the
sources for which we could not obtain a redshift may be similar to
those of 6C**0754+5019 and 6C**0935+4348, with higher or lower degrees
of Ly$\alpha$ suppression and/or chemical enrichment indicators as
dictated by their evolutionary stage (Vernet et al. 2001).

\section{Summary}\label{sec:conclusions}

\begin{itemize}

\item We have defined a sample of 68 sources from the 6C catalogue by
applying spectral index and angular size criteria to the sources with
$S_{151} > 0.5$\,Jy in a 0.421\,sr patch of sky.

\item An extensive programme of deep $K$-band imaging of all the 68
members of the 6C** sample has been presented. High-resolution VLA
radio images have been also presented for 42 of the sources. We find
\kband identifications for all but two of the sources, down to a
$3\sigma$ limiting magnitude of $K \sim 22$\,mag in a 3-arcsec aperture.

\item The distribution of \kband magnitudes of the 6C** sample is
found to be similar to that of 6C*, peaking at the same median
$K$-magnitude of 18.7~mag in a 3-arcsec aperture.
Moreover, we find that the two distributions are statistically
indistinguishable, suggesting similar redshift distributions.

\item We have also presented the results of optical spectroscopy of 27
  sources in the 6C** sample. The 15 new redshifts presented here
  together with 7 others found in the literature bring the total
  number of redshifts in the 6C** sample to 22 out of 68 (32 per cent)
  sources. The redshift content of the 6C** sample is therefore not
  complete. For the remaining sources, we will present redshift
  estimates based on detailed modelling of the $K-z$
  diagram (Paper II).

\item We find two sources without any distinctive emission or
  absorption features and nine other sources which did not yield any
  optical continuum or line emission. Some of these may be in the
  `redshift desert' region of $1.2 < z < 1.8$. Others, will have $z >
  1.8$ but feature a weak emission line spectra and/or different
  levels of Ly$\alpha$ suppression.

\item Eight of the optically identified sources are spatially compact,
implying an unresolved nuclear source. From our spectroscopy we find
that two of these sources show broad lines in their spectra,
confirming them as quasars. Two other of these sources are confirmed
as quasars in the literature, and bring the number spectroscopically
identified quasars in the 6C** sample to four.

\end{itemize}

\section*{ACKNOWLEDGMENTS} 
MJC acknowledges the support from the Portuguese Funda\c{c}\~{a}o para
a Ci\^{e}ncia e a Tecnologia, and the receipt of a Marie Curie Early
Stage Training Fellowship. MJC is also grateful to the Leiden
Observatory, where parts of this paper were written and, in
particular, to Huub R\"ottgering.  KMB acknowledges the Royal Society
for a University Research Fellowship.  CS thanks PPARC for financial
support in the form of an Advanced Fellowship.  The work of SDC was
performed under the auspices of the US Department of Energy, National
Nuclear Security Administration, by the University of California,
Lawrence Livermore National Laboratory, under contract W-7405-Eng-48.
We thank the staff at the WHT, UKIRT and Gemini for technical support.
The United Kingdom Infrared Telescope is operated by the Joint
Astronomy Centre on behalf of the U.K. Particle Physics and Astronomy
Research Council.  The William Herschel Telescope is operated on the
island of La Palma by the Isaac Newton Group in the Spanish
Observatorio del Roque de Los Muchachos of the Instituto de
Astrofisica de Canarias. This research is based on observations
obtained at the Gemini Observatory, which is operated by the
Association of Universities for Research in Astronomy, Inc., under a
cooperative agreement with the NSF on behalf of the Gemini
partnership: the National Science Foundation (United States), the
Particle Physics and Astronomy Research Council (United Kingdom), the
National Research Council (Canada), CONICYT (Chile), the Australian
Research Council (Australia), CNPq (Brazil) and CONICET (Argentina).
Data presented herein were obtained at the W. M. Keck Observatory,
which is operated as a scientific partnership among the California
Institute of Technology, the University of California, and the
National Aeronautics and Space Administration. The Observatory was
made possible by the generous financial support of the W. M. Keck
Foundation.  The VLA is part of the USA's National Radio Astronomy
Observatory which is a facility of the National Science Foundation
operated under cooperative agreement by Associated Universities, Inc.
This research has made use of the NASA/IPAC Extragalactic Database
(NED) which is operated by the Jet Propulsion Laboratory, California
Institute of Technology, under contract with the National Aeronautics
and Space Administration.  The Digitized Sky Surveys were produced at
the Space Telescope Science Institute under U.S. Government grant NAG
W-2166. The images of these surveys are based on photographic data
obtained using the Oschin Schmidt Telescope on Palomar Mountain and
the UK Schmidt Telescope. The plates were processed into the present
compressed digital form with the permission of these institutions. The
Second Palomar Observatory Sky Survey (POSS-II) was made by the
California Institute of Technology with funds from the National
Science Foundation, the National Geographic Society, the Sloan
Foundation, the Samuel Oschin Foundation, and the Eastman Kodak
Corporation.  This publication makes use of data products from the Two
Micron All Sky Survey, which is a joint project of the University of
Massachusetts and the Infrared Processing and Analysis
Center/California Institute of Technology, funded by the National
Aeronautics and Space Administration and the National Science
Foundation.

\appendix

\section[]{The survey flux densities of the 6C** sources}\label{app:a}

In this appendix we list the 6C and NVSS flux densities and positions
for all the members of the final 6C** sample
(Table~\ref{tab:radiosurvey}). The spectral indices derived from these
values are also listed.

\begin{table*}
\small
\begin{center}
\begin{tabular}{lllllccc}
\hline
\mc{1}{c}{6C Source} & \mc{2}{c}{6C Position} & \mc{2}{c}{NVSS Position} & \mc{1}{c}{6C Flux} & \mc{1}{c}{NVSS Flux} & \\
\mc{1}{c}{Name} & \mc{2}{c}{J2000} & \mc{2}{c}{J2000} & \mc{1}{c}{151\,MHz} & \mc{1}{c}{1.4\,GHz} & \mc{1}{c}{$\rm \alpha^{1400}_{151}$}\\
\mc{1}{c}{} & \mc{1}{c}{R.A.} & \mc{1}{c}{Dec.} & \mc{1}{c}{R.A.} & \mc{1}{c}{Dec.} & \mc{1}{c}{(Jy)} & \mc{1}{c}{(Jy)} & \\
\hline
6C0714+4616 &  07 17 58.3 & +46 11 24 & 07 17 58.5 & +46 11 39 & 1.65 $\pm$ 0.04 & 0.1020 $\pm$ 0.0032 & 1.25 $\pm$ 0.02\\
6C0717+5121 &  07 21 28.5 & +51 15 49 & 07 21 27.3 & +51 15 51 & 1.24 $\pm$ 0.04 & 0.1050 $\pm$ 0.0032 & 1.11 $\pm$ 0.02\\
6C0726+4938 &  07 30 05.8 & +49 32 25 & 07 30 06.1 & +49 32 41 & 0.61 $\pm$ 0.04 & 0.0432 $\pm$ 0.0014 & 1.19 $\pm$ 0.03\\
6C0737+5618 &  07 41 11.8 & +56 11 29 & 07 41 15.4 & +56 11 35 & 0.74 $\pm$ 0.04 & 0.0429 $\pm$ 0.0014 & 1.28 $\pm$ 0.03\\
6C0744+3702 &  07 47 30.3 & +36 54 42 & 07 47 29.4 & +36 54 38 & 0.64 $\pm$ 0.04 & 0.0343 $\pm$ 0.0011 & 1.31 $\pm$ 0.03\\
\\													        	
6C0746+5445 &  07 50 25.4 & +54 38 05 & 07 50 24.7 & +54 38 06 & 0.53 $\pm$ 0.04 & 0.0523 $\pm$ 0.0017 & 1.04 $\pm$ 0.04\\
6C0754+5019 &  07 58 06.3 & +50 11 01 & 07 58 06.1 & +50 11 04 & 1.05 $\pm$ 0.04 & 0.0958 $\pm$ 0.0030 & 1.07 $\pm$ 0.02\\
6C0754+4640 &  07 58 29.5 & +46 32 28 & 07 58 29.6 & +46 32 33 & 0.69 $\pm$ 0.04 & 0.0630 $\pm$ 0.0020 & 1.07 $\pm$ 0.03\\
6C0801+4903 &  08 04 40.6 & +48 54 56 & 08 04 41.3 & +48 55 00 & 1.08 $\pm$ 0.04 & 0.0871 $\pm$ 0.0027 & 1.13 $\pm$ 0.02\\
6C0810+4605 &  08 14 30.4 & +45 56 39 & 08 14 30.3 & +45 56 39 &10.26 $\pm$ 0.04 & 1.0800 $\pm$ 0.0332 & 1.01 $\pm$ 0.07\\
\\													        	
6C0813+3725 &  08 16 54.3 & +37 15 47 & 08 16 53.7 & +37 15 53 & 0.50 $\pm$ 0.04 & 0.0316 $\pm$ 0.0010 & 1.24 $\pm$ 0.04\\
6C0824+5344 &  08 27 58.7 & +53 34 20 & 08 27 58.9 & +53 34 15 & 0.88 $\pm$ 0.04 & 0.0821 $\pm$ 0.0025 & 1.06 $\pm$ 0.02\\
6C0829+3902 &  08 32 44.6 & +38 52 21 & 08 32 45.3 & +38 52 16 & 0.51 $\pm$ 0.04 & 0.0388 $\pm$ 0.0012 & 1.16 $\pm$ 0.04\\
6C0832+4420 &  08 35 26.2 & +44 09 47 & 08 35 27.5 & +44 09 52 & 0.52 $\pm$ 0.04 & 0.0412 $\pm$ 0.0013 & 1.14 $\pm$ 0.04\\
6C0832+5443 &  08 36 10.4 & +54 33 11 & 08 36 09.6 & +54 33 25 & 0.60 $\pm$ 0.04 & 0.0613 $\pm$ 0.0019 & 1.02 $\pm$ 0.03\\
\\													        	
6C0834+4129 &  08 37 48.0 & +41 19 28 & 08 37 49.2 & +41 19 54 & 0.50 $\pm$ 0.04 & 0.0539 $\pm$ 0.0017 & 1.00 $\pm$ 0.04\\
6C0848+4927 &  08 52 16.0 & +49 15 50 & 08 52 14.8 & +49 15 44 & 0.94 $\pm$ 0.04 & 0.0950 $\pm$ 0.0029 & 1.03 $\pm$ 0.02\\
6C0848+4803 &  08 52 19.1 & +47 52 04 & 08 52 17.9 & +47 52 20 & 0.71 $\pm$ 0.04 & 0.0425 $\pm$ 0.0013 & 1.26 $\pm$ 0.03\\
6C0849+4658 &  08 53 09.1 & +46 46 57 & 08 53 09.4 & +46 47 00 & 3.50 $\pm$ 0.04 & 0.3530 $\pm$ 0.0108 & 1.03 $\pm$ 0.01\\
6C0854+3500 &  08 57 13.8 & +34 48 46 & 08 57 16.0 & +34 48 24 & 0.87 $\pm$ 0.04 & 0.0812 $\pm$ 0.0025 & 1.06 $\pm$ 0.02\\
\\													        	
6C0855+4428 &  08 58 38.2 & +44 16 34 & 08 58 38.5 & +44 16 25 & 0.94 $\pm$ 0.04 & 0.0887 $\pm$ 0.0028 & 1.06 $\pm$ 0.02\\
6C0856+4313 &  08 59 19.9 & +43 02 04 & 08 59 20.1 & +43 02 01 & 0.59 $\pm$ 0.04 & 0.0630 $\pm$ 0.0020 & 1.00 $\pm$ 0.03\\
6C0902+3827 &  09 05 11.3 & +38 15 31 & 09 05 13.0 & +38 14 35 & 1.60 $\pm$ 0.04 & 0.1590 $\pm$ 0.0049 & 1.04 $\pm$ 0.02\\
6C0903+4251 &  09 06 25.2 & +42 39 04 & 09 06 26.1 & +42 39 05 & 3.14 $\pm$ 0.04 & 0.2820 $\pm$ 0.0087 & 1.08 $\pm$ 0.01\\
6C0909+4317 &  09 13 00.5 & +43 05 16 & 09 13 00.8 & +43 05 21 & 3.36 $\pm$ 0.04 & 0.3550 $\pm$ 0.0110 & 1.01 $\pm$ 0.01\\
\\													        	
6C0912+3913 &  09 16 05.8 & +39 00 35 & 09 16 05.1 & +39 00 23 & 0.56 $\pm$ 0.04 & 0.0580 $\pm$ 0.0018 & 1.02 $\pm$ 0.03\\
6C0920+5308 &  09 23 55.1 & +52 56 02 & 09 23 47.6 & +52 56 44 & 0.56 $\pm$ 0.04 & 0.0570 $\pm$ 0.0018 & 1.03 $\pm$ 0.03\\
6C0922+4216 &  09 25 59.3 & +42 03 42 & 09 25 59.6 & +42 03 36 & 2.70 $\pm$ 0.04 & 0.2570 $\pm$ 0.0079 & 1.06 $\pm$ 0.01\\
6C0924+4933 &  09 27 56.6 & +49 20 53 & 09 27 55.6 & +49 21 15 & 0.93 $\pm$ 0.04 & 0.0888 $\pm$ 0.0027 & 1.05 $\pm$ 0.02\\
6C0925+4155 &  09 28 22.6 & +41 42 21 & 09 28 22.2 & +41 42 22 & 0.91 $\pm$ 0.04 & 0.0961 $\pm$ 0.0030 & 1.01 $\pm$ 0.02\\
\\													        	
6C0928+4203 &  09 31 38.2 & +41 49 46 & 09 31 38.5 & +41 49 44 & 2.04 $\pm$ 0.04 & 0.1350 $\pm$ 0.0041 & 1.22 $\pm$ 0.02\\
6C0928+5557 &  09 32 17.3 & +55 44 36 & 09 32 17.5 & +55 44 41 & 0.58 $\pm$ 0.04 & 0.0570 $\pm$ 0.0018 & 1.04 $\pm$ 0.03\\
6C0930+4856 &  09 34 16.6 & +48 43 12 & 09 34 14.7 & +48 42 44 & 0.66 $\pm$ 0.04 & 0.0686 $\pm$ 0.0023 & 1.02 $\pm$ 0.03\\
6C0935+4348 &  09 38 19.8 & +43 34 29 & 09 38 21.4 & +43 34 37 & 1.09 $\pm$ 0.04 & 0.0523 $\pm$ 0.0021 & 1.36 $\pm$ 0.02\\
6C0935+5548 &  09 39 05.3 & +55 35 03 & 09 39 04.5 & +55 35 10 & 0.90 $\pm$ 0.04 & 0.0941 $\pm$ 0.0029 & 1.01 $\pm$ 0.02\\
\hline
\end{tabular}
\end{center}
{\caption{\label{tab:radiosurvey} The survey flux densities of the 6C** sources in Jy and the spectral indices derived from these. Spectral indices are determined by: ${\rm \alpha^{1400}_{151}} = - (\log_{10} S_{1400} - \log_{10} S_{151}) / (\log_{10} 1400 - \log_{10} 151)$. Error analysis on Parts II and III of the 6C catalogue shows that \mbox{$\sim$ 60 per cent} of the sources have flux densities within $\pm$ 40 mJy of their true value (Hales et al. 1988, 1990) - thus we take 40 mJy as the typical error for the 6C flux densities quoted here.
}}
\end{table*}

\addtocounter{table}{-1}
\begin{table*}
\small
\begin{center}
\begin{tabular}{lllllccc}
\hline
\mc{1}{c}{6C Source} & \mc{2}{c}{6C Position} & \mc{2}{c}{NVSS Position} & \mc{1}{c}{6C Flux} & \mc{1}{c}{NVSS Flux} & \\
\mc{1}{c}{Name} & \mc{2}{c}{J2000} & \mc{2}{c}{J2000} & \mc{1}{c}{151\,MHz} & \mc{1}{c}{1.4\,GHz} & \mc{1}{c}{$\rm \alpha^{1400}_{151}$}\\
\mc{1}{c}{} & \mc{1}{c}{R.A.} & \mc{1}{c}{Dec.} & \mc{1}{c}{R.A.} & \mc{1}{c}{Dec.} & \mc{1}{c}{(Jy)} & \mc{1}{c}{(Jy)} & \\
\hline
6C0938+3801 &  09 41 51.6 & +37 47 43 & 09 41 52.4 & +37 47 22 & 1.03 $\pm$ 0.04 & 0.0836 $\pm$ 0.0027 & 1.13 $\pm$ 0.02\\
6C0943+4034 &  09 46 27.2 & +40 20 29 & 09 46 27.3 & +40 20 32 & 0.99 $\pm$ 0.04 & 0.0929 $\pm$ 0.0029 & 1.06 $\pm$ 0.02\\
6C0944+3946 &  09 47 47.8 & +39 32 44 & 09 47 49.1 & +39 33 11 & 0.66 $\pm$ 0.04 & 0.0705 $\pm$ 0.0022 & 1.00 $\pm$ 0.03\\
6C0956+4735 &  09 59 18.8 & +47 21 18 & 09 59 18.7 & +47 21 14 & 6.13 $\pm$ 0.04 & 0.4930 $\pm$ 0.0152 & 1.13 $\pm$ 0.01\\
6C0957+3955 &  10 00 46.9 & +39 41 01 & 10 00 46.1 & +39 40 45 & 0.62 $\pm$ 0.04 & 0.0650 $\pm$ 0.0020 & 1.01 $\pm$ 0.03\\
\\								     					        	
6C1003+4827 &  10 06 40.9 & +48 13 11 & 10 06 40.5 & +48 13 09 & 6.88 $\pm$ 0.04 & 0.6030 $\pm$ 0.0186 & 1.09 $\pm$ 0.01\\
6C1004+4531 &  10 07 40.7 & +45 16 27 & 10 07 42.9 & +45 16 07 & 0.70 $\pm$ 0.04 & 0.0738 $\pm$ 0.0023 & 1.01 $\pm$ 0.03\\
6C1006+4135 &  10 09 29.5 & +41 21 14 & 10 09 27.5 & +41 20 46 & 0.52 $\pm$ 0.04 & 0.0545 $\pm$ 0.0017 & 1.01 $\pm$ 0.04\\
6C1009+4327 &  10 12 10.2 & +43 13 09 & 10 12 09.7 & +43 13 06 & 2.89 $\pm$ 0.04 & 0.1880 $\pm$ 0.0058 & 1.23 $\pm$ 0.01\\
6C1015+5334 &  10 18 29.8 & +53 19 26 & 10 18 30.0 & +53 19 34 & 1.44 $\pm$ 0.04 & 0.1400 $\pm$ 0.0043 & 1.05 $\pm$ 0.02\\
\\								     					        	
6C1017+3436 &  10 20 05.8 & +34 21 26 & 10 20 05.7 & +34 21 21 & 1.17 $\pm$ 0.04 & 0.1150 $\pm$ 0.0035 & 1.04 $\pm$ 0.02\\
6C1018+4000 &  10 21 29.0 & +39 45 24 & 10 21 28.6 & +39 45 46 & 0.53 $\pm$ 0.04 & 0.0546 $\pm$ 0.0017 & 1.02 $\pm$ 0.04\\
6C1035+4245 &  10 38 40.7 & +42 29 46 & 10 38 41.0 & +42 29 51 & 1.89 $\pm$ 0.04 & 0.1080 $\pm$ 0.0033 & 1.28 $\pm$ 0.02\\
6C1036+4721 &  10 39 15.6 & +47 05 37 & 10 39 15.7 & +47 05 40 & 3.70 $\pm$ 0.04 & 0.3700 $\pm$ 0.0114 & 1.03 $\pm$ 0.01\\
6C1043+3714 &  10 46 11.6 & +36 58 25 & 10 46 11.9 & +36 58 45 & 2.62 $\pm$ 0.04 & 0.2590 $\pm$ 0.0079 & 1.04 $\pm$ 0.01\\
\\								     					        	
6C1044+4938 &  10 47 47.4 & +49 22 40 & 10 47 47.9 & +49 22 36 & 1.66 $\pm$ 0.04 & 0.1490 $\pm$ 0.0046 & 1.08 $\pm$ 0.02\\
6C1045+4459 &  10 48 31.3 & +44 44 06 & 10 48 32.2 & +44 44 27 & 0.95 $\pm$ 0.04 & 0.0809 $\pm$ 0.0025 & 1.11 $\pm$ 0.02\\
6C1048+4434 &  10 51 26.1 & +44 18 18 & 10 51 26.5 & +44 18 22 & 1.51 $\pm$ 0.04 & 0.1540 $\pm$ 0.0047 & 1.02 $\pm$ 0.02\\
6C1050+5440 &  10 53 36.0 & +54 24 36 & 10 53 36.3 & +54 24 42 & 0.93 $\pm$ 0.04 & 0.0647 $\pm$ 0.0020 & 1.20 $\pm$ 0.02\\
\\								     					        	
6C1052+4349 &  10 55 38.3 & +43 33 17 & 10 55 37.4 & +43 33 36 & 0.51 $\pm$ 0.04 & 0.0510 $\pm$ 0.0016 & 1.03 $\pm$ 0.04\\
6C1056+5730 &  10 59 14.9 & +57 14 46 & 10 59 15.0 & +57 14 45 & 2.66 $\pm$ 0.04 & 0.2190 $\pm$ 0.0068 & 1.12 $\pm$ 0.01\\
6C1100+4417 &  11 03 34.5 & +44 01 10 & 11 03 33.5 & +44 01 26 & 0.72 $\pm$ 0.04 & 0.0639 $\pm$ 0.0020 & 1.09 $\pm$ 0.03\\
6C1102+4329 &  11 05 42.6 & +43 13 21 & 11 05 42.9 & +43 13 24 & 1.11 $\pm$ 0.04 & 0.0990 $\pm$ 0.0031 & 1.08 $\pm$ 0.02\\
\\								     					        	
6C1103+5352 & 11 06 14.9  & +53 35 57 & 11 06 14.9 & +53 36 00 & 2.67 $\pm$ 0.04 & 0.2700 $\pm$ 0.0082 & 1.03 $\pm$ 0.01\\ 
6C1105+4454 &  11 08 45.6 & +44 38 16 & 11 08 46.1 & +44 38 14 & 0.83 $\pm$ 0.04 & 0.0874 $\pm$ 0.0028 & 1.01 $\pm$ 0.03\\
6C1106+5301 &  11 09 47.9 & +52 45 20 & 11 09 48.9 & +52 45 17 & 0.77 $\pm$ 0.04 & 0.0617 $\pm$ 0.0019 & 1.13 $\pm$ 0.03\\
6C1112+4133 &  11 15 11.2 & +41 17 36 & 11 15 09.8 & +41 17 02 & 0.54 $\pm$ 0.04 & 0.0282 $\pm$ 0.0009 & 1.33 $\pm$ 0.04\\
6C1125+5548 &  11 28 30.7 & +55 31 30 & 11 28 26.9 & +55 33 11 & 0.63 $\pm$ 0.04 & 0.0404 $\pm$ 0.0013 & 1.23 $\pm$ 0.03\\
\\								     					        	
6C1132+3209 &  11 35 26.2 & +31 53 11 & 11 35 26.7 & +31 53 33 & 0.63 $\pm$ 0.04 & 0.0622 $\pm$ 0.0020 & 1.04 $\pm$ 0.03\\
6C1135+5122 &  11 38 26.2 & +51 06 07 & 11 38 27.8 & +51 05 56 & 0.66 $\pm$ 0.04 & 0.0570 $\pm$ 0.0018 & 1.10 $\pm$ 0.03\\
6C1138+3309 &  11 41 23.7 & +32 52 58 & 11 41 25.8 & +32 52 14 & 0.93 $\pm$ 0.04 & 0.0618 $\pm$ 0.0019 & 1.22 $\pm$ 0.02\\
6C1138+3803 &  11 41 29.1 & +37 46 54 & 11 41 30.4 & +37 46 54 & 0.51 $\pm$ 0.04 & 0.0487 $\pm$ 0.0015 & 1.05 $\pm$ 0.04\\
6C1149+3509 &  11 51 48.7 & +34 53 00 & 11 51 50.6 & +34 53 02 & 0.61 $\pm$ 0.04 & 0.0576 $\pm$ 0.0018 & 1.06 $\pm$ 0.03\\
\hline
\end{tabular}
\end{center}
{\caption{\label{tab:radiosurvey} {\em continued}
}}
\end{table*}

\end{document}